\documentclass{article}
\usepackage{emulateapj}
\usepackage{float,epsfig,psfig}

\newcommand{\LA}{\mbox{\raisebox{-0.6ex}{$\stackrel{\textstyle<}{\sim}$}}}
\newcommand{\GA}{\mbox{\raisebox{-0.6ex}{$\stackrel{\textstyle>}{\sim}$}}}

\newcommand{\cha}{{\sl Chandra}}
\newcommand{\ein}{{\sl Einstein}}
\newcommand{\hst}{{\sl Hubble}}
\newcommand{\msun}{M$_{\odot}$}
\newcommand{\ergl}{erg~s$^{-1}$}

\newcommand{\mdot}{$\dot{M}$}
\newcommand{\ros}{{\sl ROSAT}}
\newcommand{\asca}{{\sl ASCA}}
\newcommand{\etal}{et al.}
\newcommand{\sss}{supersoft source}
\newcommand{\ssss}{supersoft sources}

\newcommand{\Sss}{Supersoft sources}

\slugcomment{Submitted to Astrophysical Journal}

\begin{document}

\title{
{\sl Chandra} Discovery of Luminous Supersoft X-ray Sources in M81}

\author{
Douglas~A.~Swartz\altaffilmark{1},
Kajal~K. Ghosh\altaffilmark{1},
Valery~Suleimanov\altaffilmark{2},
Allyn~F.~Tennant\altaffilmark{3},
Kinwah~Wu\altaffilmark{4}
} 
\altaffiltext{1}{Universities Space Research Association, 
NASA Marshall Space Flight Center, SD50, Huntsville, AL, USA}
\altaffiltext{2}{Kazan State University, Kremlevskaya str. 18, 
420008 Kazan, Russia}
\altaffiltext{3}{Space Science Department, 
NASA Marshall Space Flight Center, SD50, Huntsville, AL, USA} 
\altaffiltext{4}{MSSL, University College London, Holmbury St. Mary, Surrey, RH5 6NT, UK; and School of Physics, University of Sydney, NSW 2006, Australia}

\begin{abstract}
A \cha\ ACIS-S 
imaging observation of the nearby galaxy M81 (NGC 3031)
reveals 9 luminous soft X-ray sources.
The local environments, X-ray spectral properties, and X-ray 
light curves of the sources
are presented and discussed in the context of prevailing 
physical models for \ssss.
It is shown that the sample falls within expectations based on
population synthesis models taken from the literature though
the high observed luminosities ($L_{obs} \sim 2 \times 10^{36}$ to
$\sim 3 \times 10^{38}$ erg-s$^{-1}$ in the 0.2--2.0~keV band) 
and equivalent blackbody temperatures 
($T_{eff} \sim 40$ to 80~eV) 
place the brightest detected M81 objects at the high luminosity end of
the class of \ssss\ defined by previous \ros\ and \ein\ 
studies of nearby galaxies.
This is interpreted as a natural consequence of the 
higher sensitivity of \cha\ to 
hotter and more luminous systems.
Most of the sources can be explained as canonical \ssss,
accreting white dwarfs powered by steady surface nuclear burning,
with X-ray spectra well-fit by hot white dwarf 
local thermodynamic equilibrium atmosphere models.
An exceptionally bright source is scrutinized in greater detail as
its estimated bolometric luminosity, 
$L_{bol} \sim 1.5 \times 10^{39}$ \ergl,
greatly exceeds 
theoretical estimates for \ssss.
This source may be beyond the 
stability limit and undergoing a phase of mass outflow under extreme conditions.
Alternatively, a model in which 
the observed X-ray spectrum arises from an accretion disk 
around a blackhole of mass $\sim$1200/(cos$i)^{1/2}$ \msun\
(viewed at an inclination angle $i$) cannot be excluded.
\end{abstract}


\section{Introduction}

Luminous supersoft X-ray sources have
effective blackbody temperatures of 15--80~eV and bolometric luminosities
$10^{36}$ to $10^{38}$~erg~s$^{-1}$ (Kahabka \& van den Heuvel 1997). 
Lacking significant emission above $\sim$0.5--1~keV, \ssss\ are easily 
identified
using even low-resolution X-ray detectors.
However,
only 10 Galactic objects are listed as \ssss\ 
in the latest compilation of Greiner (2000a) while
population synthesis studies 
(Di~Stefano \& Rappaport 1994, Yungleson \etal\ 1996, 
Di~Stefano \& Nelson 1996) predict 
of order 1000 \ssss\ could be active in ordinary
galaxies like our own.
Local objects are therefore of limited use for \sss\
population studies. The best opportunity is to explore
nearby galaxies at high galactic latitude 
where intervening absorption is low, all
the sources are at a common distance, 
a single pointing can detect a large number of sources,
and their spatial distribution can be measured.
Indeed, surveys conducted using the \ros\ and \ein\ 
observatories have identified 
8 \ssss\ in the LMC, 4 in the SMC, and 34 in M31 
(see Greiner 1996 for a detailed listing). 
The high sensitivity and moderate spectral resolution 
make the \cha\ X-ray Observatory an ideal 
facility to systematically extend \sss\ population 
studies to other nearby galaxies.
This is the rationale for the present investigation.

A 50~ks \cha\ ACIS-S imaging observation 
of the nearby galaxy M81 reveals
nine \sss\ candidates 
identified by broadband X-ray colors
(\S\ref{s:observations}).
Four of these sources are located within the old 
population of bulge stars and all but one of 
the remainder are coincident with M81 spiral arms. 
A search of known objects finds
two of the bulge sources are spatially coincident with sources of enhanced
[OIII] $\lambda$5007~\AA\ emission; 
consistent with nebulae photoionized by the X-ray sources.
No other spatial correlations were discovered in searches 
of multiwavelength archival data, 
contemporary radio and H$\alpha$ images,
or published source catalogues. 
The two brightest \sss\ candidates are visible in 
archival \ros\ PSPC or HRI X-ray images.

Detailed X-ray spectral analysis
of the three brightest sources is presented in 
\S\ref{s:spectral_analysis}.
Their effective temperatures, based on
line-blanketed local thermodynamic equilibrium white dwarf atmospheres and 
blackbody model fits, span 
$T_{eff}$$\sim$$50$ to 80~eV.
The corresponding
inferred bolometric luminosities are 
$L_{bol}$$\sim$$3\times 10^{38}$ to $\sim$$1.5 \times 10^{39}$~\ergl\
ranking the brightest M81 objects among the hottest and 
brightest of all previously-known
\ssss.
Extrapolating the spectral fits to the fainter sources
implies the observed luminosities of the M81 \sss\ population
range from 
$L_{obs}$$\sim$$2 \times 10^{36}$ (near the sensitivity limit) to
$\sim$$3 \times 10^{38}$ erg-s$^{-1}$ in the 0.2--2.0~keV band.
The brightest sources show signs of X-ray variability over the duration of the
\cha\ observation and over the approximately 10-year light 
curve observed by \ros\ 
(\S\ref{s:lc_analysis}).

The X-ray signatures of the M81 sources are 
compared to well-studied objects in the (rather heterogeneous) class of \ssss\
in an effort to identify possible physical mechanisms responsible 
for the observed emission
(\S\ref{s:discussion}).
The brightest source cannot easily be explained by prevailing 
theory while the other two sources for which reliable spectral fits 
can be made show strong similarities to the
hot close-binary accreting white dwarf (WD)
systems CAL~87 and CAL~83, 
respectively.
Some prospects for future observations are discussed (\S\ref{s:conclusions})
and a comparison is made  
to the \sss\ populations observed in M31 with \ros\ 
(Supper \etal\ 1997, Kahabka 1999) 
and in M101 by \cha\ (Pence \etal\ 2001). 

\section{Observation of Supersoft Sources in M81} \label{s:observations}

A 49926 second observation of a portion of the galaxy M81 was 
obtained with the \cha\ Advanced CCD Imaging Spectrometer (ACIS) 
on 2000 May 7.
The nucleus and bulge of the galaxy were located near the 
center of the back-illuminated S3 device of the spectroscopy 
array operating in imaging mode.
The observation was taken in faint timed exposure mode at 
3.241 s-frame$^{-1}$.
The global properties of the sources detected in this data 
are given in Tennant \etal\ (2001).
Standard \cha\ X-ray Center processing has applied aspect 
corrections and compensated for spacecraft dither. 
A charge transfer inefficiency (CTI) corrector algorithm 
(Townsley \etal\ 2000)\footnote[5]{Matching response matrices were also provided by L. Townsley}
was then applied to the Level~1 event list to partially correct for 
the charge loss and charge smearing effects of CTI in the ACIS detectors.
The data were then cleaned of bad pixels and columns.
We selected the standard grade set and events in pulse 
invariant (PI) channels corresponding to $\sim 0.1$ to $8.0$ keV 
for source detection (\S~\ref{s:selection}) 
and $\sim$0.2--2.0~keV for spectral analysis (\S~\ref{s:spectral_analysis}).
The 50~ks exposure corresponds to a limiting \sss\ observed 0.2--2.0~keV 
luminosity of $\sim 2 \times 10^{36}$ ergs-s$^{-1}$ 
for a 3$\sigma$ signal-to-noise ratio assuming the weak \sss\ candidates
have the same spectrum as the brightest object in the class.

The detector viewing area covers 57\% of the 
optical extent of the galaxy, 
defined as the ellipse of major diameter 26.$\arcmin$9 corresponding to the 
$D_{25}$ diameter as tabulated in de~Vaucouleurs \etal\ (1991),
oriented at position angle 157$^{\circ}$,
and with major-to-minor axis ratio 2:1 corresponding to the 60$^{\circ}$
inclination angle of M81. This area includes all of the back-illuminated (BI)
S3 device, approximately half of each of the front-illuminated 
(FI) S2 and S4 devices, and the outer corner of the FI device I3.
The data from each device is analyzed independently due to differing energy
resolutions, low-energy responses, and background signals.

In addition to this primary dataset, a 2.4-ks ACIS-S image taken 2000 Mar 21
and numerous \ros\ PSPC and HRI datasets were used to construct 
long-term light curves of the brightest 
sources (see Immler \& Wang, 2001 for details of the \ros\ observations).

\subsection{Source Identification Criteria} \label{s:selection}

X-ray sources were identified using the source detection method described in 
Tennant \etal\ (2001)
that compares the data to an analytic function that approximately 
matches the telescope's 
point-spread function (PSF) and includes off-axis broadening.
Small spatial regions 
centered on each source with a size encompassing the 95\% encircled energy 
radii of the PSF (at 1.5~keV) were then selected for analysis.
These regions are $<2\arcsec$ diameter on-axis and exceed 10$\arcsec$
only for sources \GA$8\arcmin$ off-axis.

Background-subtracted source counts were 
binned into three broad bands defined as $S$ (0.3--1.0~keV), 
$M$ (1.0--2.0~keV), and $H$ (2.0--8.0~keV) and the X-ray colors 
$MS \equiv (M-S)/(M+S)$ and $HS \equiv (H-S)/(H+S)$ were constructed.
For this purpose,
two background spectra were defined for sources on S3: 
One for the bulge where contributions from the wings of the PSF
of the bright nucleus and from a relatively high level of unresolved
X-ray emission was previously detected in the data (Tennant \etal\ 2001) 
and another for the disk. 
Separate background spectra are defined for sources on S2 and on S4. 
In all cases, regions enclosing all identified sources are first excluded from
the data then the spectrum of the remaining area was 
assigned to the background.
Figure~\ref{f:color-color} shows the resulting 
background-corrected  color-color 
diagram for all 81 sources detected on S3 above a signal-to-noise ratio of 3.5
(Tennant \etal\ 2001). The bulge and disk background spectra are 
also represented. 
Some faint sources far from the aimpoint are background-dominated.
Their background-subtracted count rates in certain energy bands 
can be negative and the absolute values of their 
resulting colors may exceed unity, as seen in Figure~\ref{f:color-color}.

\begin{center}
\includegraphics[angle=90,width=\columnwidth]{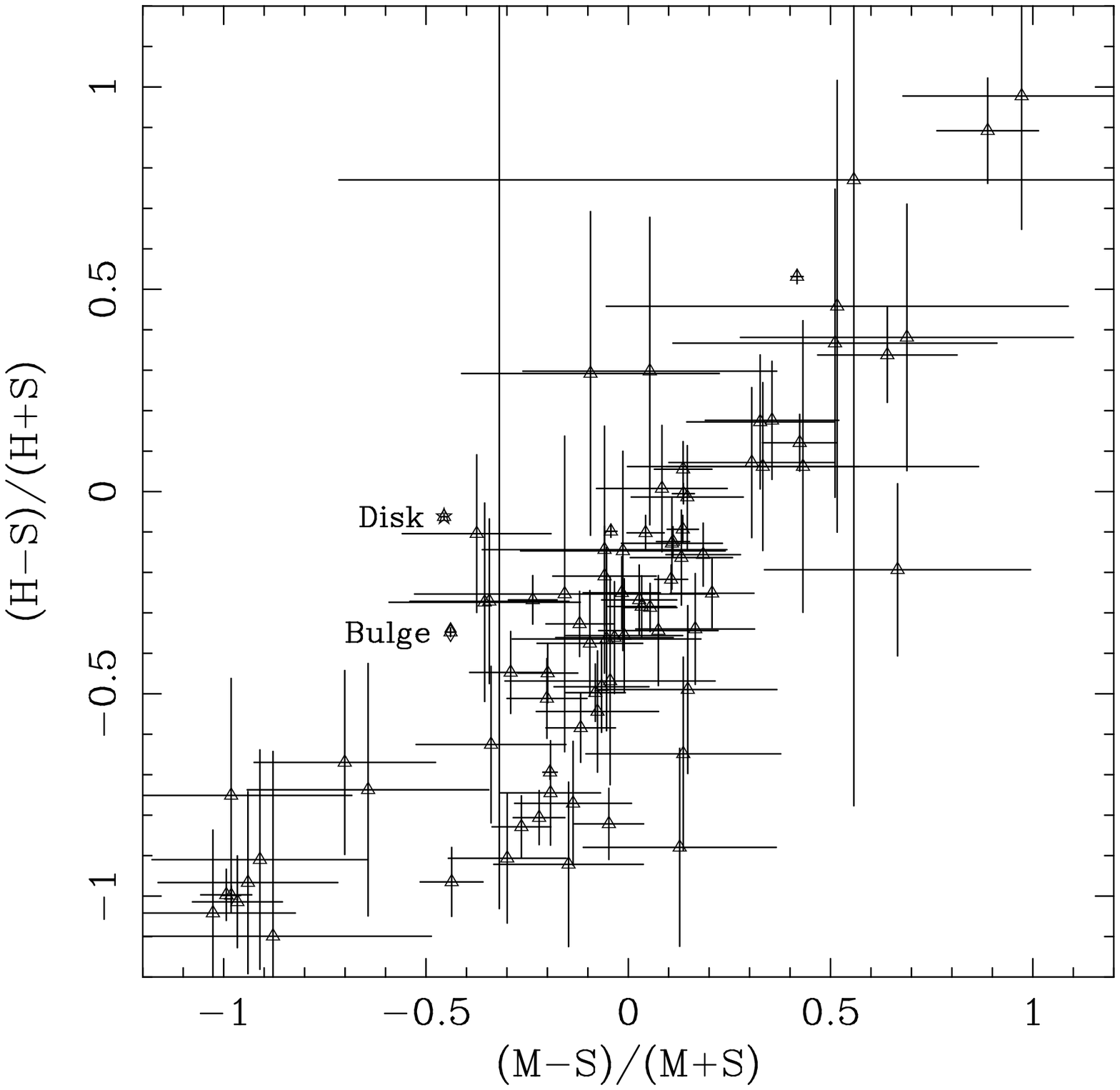}
\vspace{10pt}
\figcaption{X-ray color-color diagram for M81 sources on the BI device S3.
Background-subtracted source counts are binned into three bands;
$S$ (0.3--1.0~keV), 
$M$ (1.0--2.0~keV), and $H$ (2.0--8.0~keV).
Points labeled {\em Disk} and {\em Bulge} represent the colors of the
background spectra in these two regions. \label{f:color-color}}
\end{center}

\Sss\ lie near $(-1,-1)$ in this figure (no emission above 1.0~keV). 
Those sources with color values $(MS<-0.5,HS<-0.5)$ were tenatively
classified as supersoft sources.
This criteria selects 10 sources on S3, two sources on S2, and none on
the remaining CCDs. 

The list of sources was compared to catalogues of objects to eliminate
those with properties inconsistent with \ssss.
Supernova remnant candidate no.~17 in the catalogue of 
Matonick \& Fesen (1997) lies within the positional uncertainty
of one of the \sss\ candidates detected on S3. Though this source
is weak, its spectrum appears more similar to the other 
X-ray sources associated 
with supernova remants 
than to the other \sss\ candidates. 
Matonick \& Fesen (1997) report an [OIII]/H$\alpha$ ratio for the 
supernova remnant candidate of 0.6 and an [SII]/H$\alpha$ ratio of 1.2.
Rappaport \etal\ (1994) predict far different ratios, 2.4~--~6.1 and
0.1~--~0.5, respectively, for ionized regions surrounding \ssss.
This source is therefore 
excluded from the list of candidate \ssss\ even though its colors 
($MS=-0.88\pm0.39,HS=-1.10\pm0.45$) lie within
our selection criteria.

Both candidate sources on S2 are coincident with bright 
($m_V$$\sim$9 mag) G0 stars in the PPM catalogue of stars. While they
have X-ray colors $(MS,HS)\sim(-0.6,-0.8)$ they are
excluded from further consideration.

The positions of the remaining
candidate \ssss\ were compared to optical features in 
Digitized Sky Survey (DSS) and archival
\hst/WFPC2 images to search for potential uncatalogued
foreground objects. 
Late-type stars with active coronae and AM~Her-type magnetic 
cataclysmic variables (CVs) are the only 
foreground objects with X-ray colors falling within our 
selection criteria.
Foreground late-type stars are easily identified in
optical images as their
X-ray flux is typically a small fraction of their optical flux.
Am~Her-type mCVs 
have a soft component ($T \sim 10$--100~eV) that may 
dominate any hard bremsstrahlung
($kT \sim 10$~keV) component and appear supersoft. 
However, their K or M star companions
should be visible optically out to distances well beyond 
the Galactic disk in the direction of M81.
It remains possible that some of 
the weakest \sss\ candidates could be 
very bright CVs, with late M-type companions, and escape 
optical detection provided they are located some $\sim$7~kpc
above the plane of the Galaxy.

\begin{center}
\includegraphics[angle=90,width=\columnwidth]{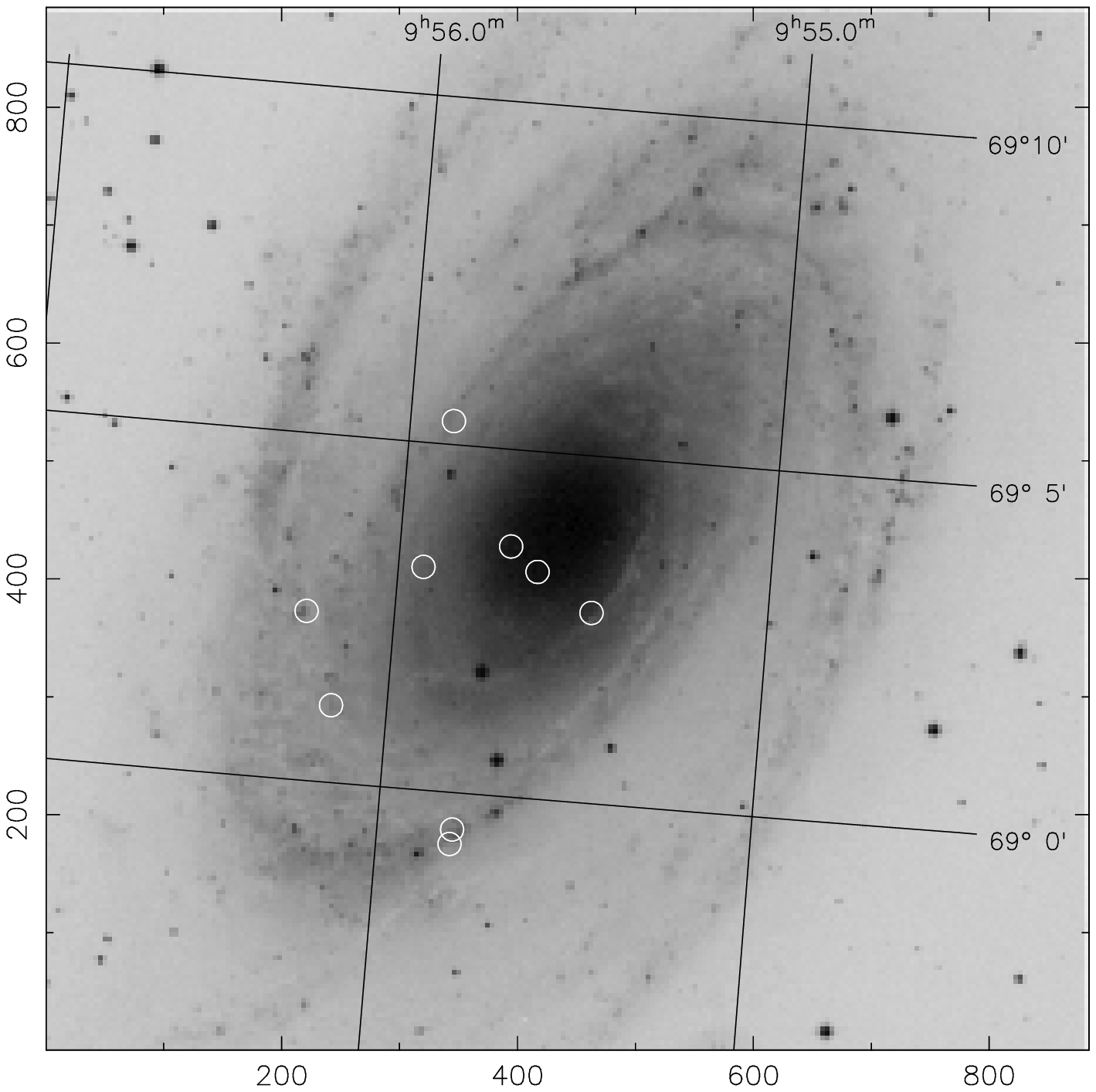}
\figcaption{Digitized Sky Survey image of M81 with the positions of the
9 X-ray-detected \sss\ candidates superposed. 
Image is $\sim$15$\times$15 arcmin. Circles have 10$\arcsec$ radii.
\label{f:full-dss-image}}
\end{center}

The observed properties of the remaining 9 
candidate supersoft sources, all located on the BI S3 device, are listed in 
Table~1 and
their positions are shown superimposed 
on the second generation DSS image of M81 in
Figure~\ref{f:full-dss-image}.
Hereafter, individual supersoft sources in M81 will be referred to in order
of observed brightness, column~1 of Table~1, 
as source N1 through source N9.

\begin{center}
\small{
\begin{tabular}{lccrl} 
\multicolumn{5}{c}{{\sc Table 1}} \\
\multicolumn{5}{c}{{\sc Observed \ssss\ in M81}} \\ 
\hline \hline
No. & RA & DEC & \multicolumn{1}{c}{Flux$^a$} 
& \multicolumn{1}{c}{location}  \\   
    & (2000) & (2000) &  &  \\
\hline
   1   & 09 55 42.15  & 69 03 36.2 & 74.95 & bulge \\ 	
   2   & 09 56 08.96  & 69 01 06.6 & 10.24 & arm   \\ 	
   3   & 09 55 53.00  & 69 05 20.3 &  3.30 & arm   \\ 	
   4   & 09 55 28.38  & 69 02 44.6 &  1.09 & bulge \\ 	
   5   & 09 55 37.58  & 69 03 16.2 &  0.95 & bulge \\ 	
   6   & 09 55 55.97  & 69 03 12.5 &  0.78 & bulge \\ 	
   7   & 09 56 14.16  & 69 02 26.2 &  0.63 & disk  \\ 	
   8   & 09 55 47.92  & 68 59 28.2 &  0.59 & arm   \\ 	
   9   & 09 55 48.13  & 68 59 15.5 &  0.41 & arm   \\ 	
%
\hline
\multicolumn{5}{l}{$^a$background-subtracted, 0.2-2.0~keV,} \\
\multicolumn{5}{l}{\hspace{5pt} in units of $10^{-3}$ cts-s$^{-1}$} \\
\label{t:observed-properties}
\end{tabular}
} 
\end{center}

The lack of candidate 
\ssss\ on the FI devices is attributable to the 
small fraction of M81 falling within
the viewing field of the FI devices, to the scarcity of X-ray sources
(of all types) found far from the central regions of M81, 
to the weaker low energy response of the FI devices,
and to the decreased off-axis source sensitivity. 
The detection limit for \ssss\ on the FI devices is 
$\sim 1.2\times 10^{37}$~\ergl\ or about 6 times brighter than the 
corresponding limit on the BI device S3.

\subsection{Local Source Environments} \label{s:environments}

None of the candidate \ssss\ appear extended in the \cha\ data. 
As listed in Table~1, candidate sources are found throughout 
the bulge and disk of M81 and several lie within 
$\sim$200~pc of spiral arms ($1\arcmin = 1.086$~kpc 
for the adopted distance of 3.6~Mpc to M81, Freedman \etal\ 1994). 

While several of the candidate \ssss\ are in crowded, 
source-rich regions of this well-studied galaxy,
only two can be unequivically identified with catalogued objects 
observed at other wavelengths
based on positional coincidence. 
These are the bulge sources N4 and N5.
Three of the 4 bulge \sss\ candidates lie within the 
$\sim$4$\arcmin$ observing field of 
the [OIII] $\lambda$5007~\AA\ survey of M81 
planetary nebulae (PNe) performed by Jacoby \etal\ (1989).
Source N4 is coincident with object ID~68 
(apparent magnitude 25.33 at $\lambda$5007~\AA) and
source N5 is coincident with object ID~116 
(apparent magnitude 24.78) in the list of Jacoby \etal\ (1989).
The observed X-ray luminosities of these sources exceeds 
that of typical planetary nebulae 
($10^{30}$ to $10^{32}$~erg~s$^{-1}$) 
by several orders of magnitude assuming they are at the
distance of M81. 
The one known exception is 1E0056.8-7154, a \sss\ associated with the 
Small Magellenic Cloud PNe N67 (Wang 1991) with a bolometric luminosity 
$2 \times 10^{37}$~\ergl\ (Heise, van~Teeseling, \& Kahabka 1994).
The lack of detectable emission at other wavelengths, 
particularly in \hst/WFPC2 images, 
indicates they are not foreground objects lying, by chance, 
along the line of sight.
Instead,
the [OIII] $\lambda$5007~\AA\ emission could come from 
regions ionized by the X-ray source in a way
similar to that observed in the LMC 
\sss\ CAL~83 (Remillard, Rappaport, \& Macri 1995). The 
flux observed by Jacoby \etal\ (1989) is
consistent with the \sss\ ionization model predictions 
of Rappaport \etal\ (1994). This is discussed in more detail in
\S\ref{s:s4ands5}.

There are no point-like objects observed at 
other wavelengths spatially coincident with any of the remaining
\sss\ candidates in M81. 
It is noteworthy that the exceptionally X-ray bright source N1 
is within the field observed by Jacoby \etal\ (1989),
and the recent survey of Magini \etal\ 2001, but is not seen at 
[OIII] $\lambda$5007~\AA.
This source lies in a featureless region of the bulge dominated in
optical bands by a high density of unresolved bulge stars. 
There is no nearby object 
discernable in archival \hst/WFPC2 images of the region,
nor in UV images centered at $\lambda$2490~\AA\ and 
$\lambda$1520~\AA\ (Hill \etal\ 1992),
a recent (2001 May 21; A. Shafter, private communication) 
CCD image taken through a 70\AA\ 
wide H$\alpha$ filter, 
the continuum-subtracted H$\alpha$ data of Devereux \etal\ (1995),
the $\lambda$6 and $\lambda$20~cm radio continuum observations 
reported by Kaufman \etal\ (1996), or recent $\lambda$6~cm data 
(T. Pannuti 2001, private communication).

\section{X-ray Spectral Analysis} \label{s:spectral_analysis}

The spectra of all 9 candidate sources 
are shown in Figure~\ref{f:all_spectra}. 
They typically show a rise 
from low energies to a peak at $\sim$0.45 to $\sim$0.7~keV 
followed by a rapid decline toward higher energies. 
The shape of the low-energy portion of the spectra is determined by
a combination of extinction
and a decreasing 
ACIS detection efficiency at low energies. 
Above the peak, the shape is determined
by a roughly exponential decline of the intrinsic source spectra.

In spectra such as these, the inferred bolometric luminosity is 
much higher than that observed with the peak of the intrinsic 
spectral energy distribution
occuring at energies lower than observed. 
The inferred luminosities are very sensitive to
the model parameters and
to calibration uncertainties in the low-energy response of the BI devices.
Though there is often significant flux below 0.2~keV, 
data at energies below this value were excluded from the fitting
because of these uncertainties.

A sufficient number of counts for spectral analysis
were accumulated from only
the three brightest sources listed in Table~1.
The spatial regions selected for spectral analysis were chosen
centered on each source as described above (\S\ref{s:selection}). 
Accompanying background spectra were extracted for each source 
from annular regions centered on the
source with inner and outer radii of 30 and 50 pixels, respectively,
and containing of order 200 counts. 
Note this procedure differs from the
large background regions selected for constructing the broadband colors 
described previously. The procedure used in this section more accurately 
represents local variations in background.
Spectra have been grouped into spectral bins containing a minimum of 20 counts
and fitted using the XSPEC spectral fitting package (Arnaud 1996) 
to blackbody and to model atmosphere spectra of hot white dwarfs 
calculated in the local thermodynamic equilibrium (LTE) approximation. 

The X-ray emission from \ssss\ is believed to originate 
from nuclear burning on the surface of a WD star.
The nuclear burning takes place at high optical depth 
within a geometrically thin, high density atmosphere 
with the resulting emergent spectrum highly modified by 
the cooler outer layers.
Under these conditions, hot WD LTE model atmosphere spectra  
is a more appropriate approximation than is a simple blackbody assumption. 
Therefore, we developed tables of plane-parallel LTE model 
atmospheres as an alternative model for spectral fitting.
Under certain circumstances
the WD atmosphere may become extended or even partially ejected
as a result of rapid nuclear burning.
Spectra of extended, expanding,
spherical atmospheres tend to be flatter than that from
plane-parallel hydrostatic atmospheres
with the same color temperature at maximum flux.
Model atmospheres including an optically thick wind are 
currently in preparation.

\begin{center}
\includegraphics[angle=90,width=\columnwidth]{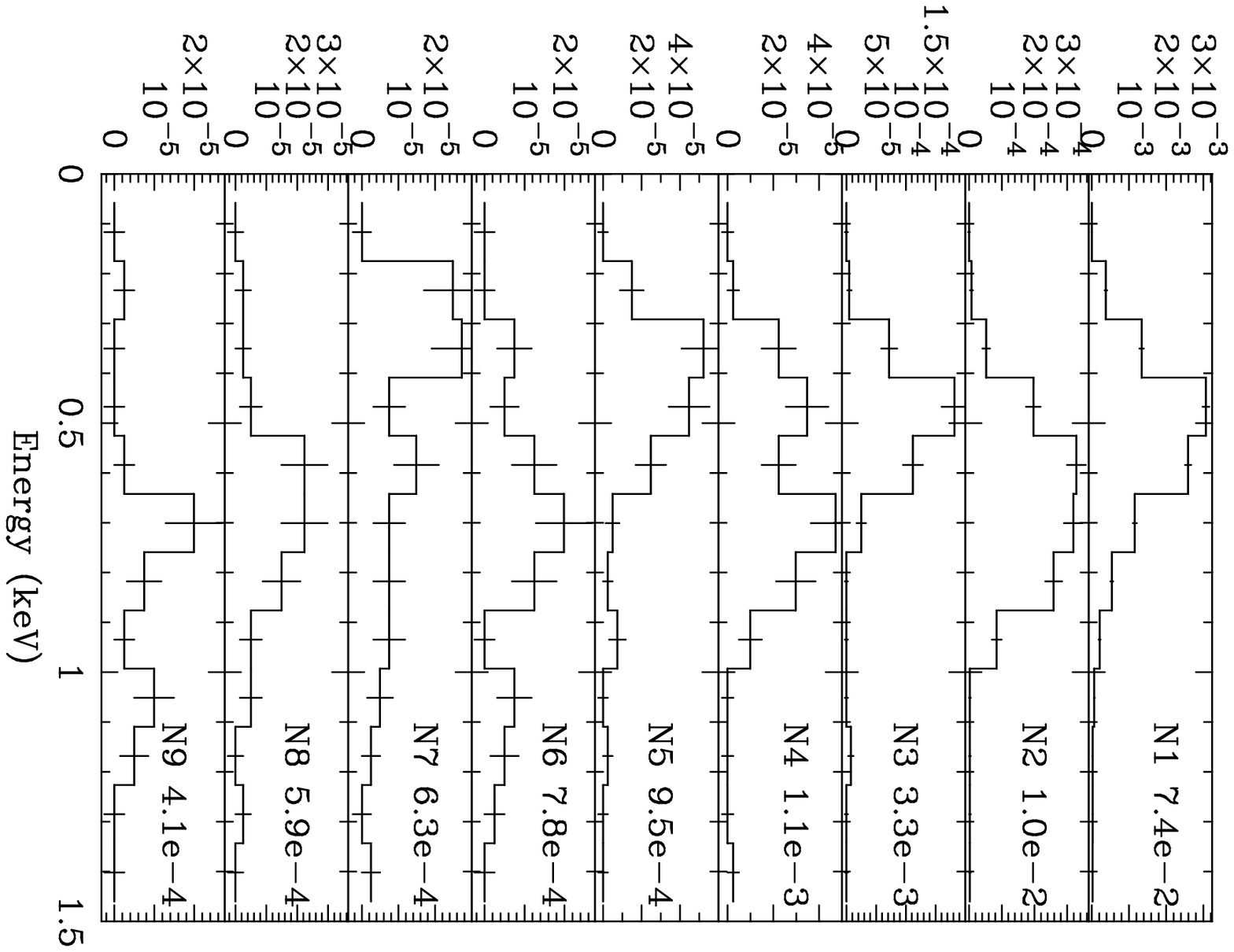}
\vspace{10pt}
\figcaption{Spectra of nine \sss\ candidates. Labels denote source number
and count rates over the 0.2--2.0~keV \cha\ energy band (Table~1).
Data have been binned by 8 PI channels ($\sim 117$~eV). 
Vertical axis units are counts/sec.
\label{f:all_spectra}}
\vspace{20pt}
\end{center}

The fundamental parameters of our plane-parallel LTE model atmospheres 
are the effective temperature, $T_{eff}$, surface gravity, $g$, and 
luminosity scale factor 
$K = L/(4\pi \sigma T_{eff}^4 D^2)$ 
where $L$ is the bolometric luminosity,
$\sigma$ is the Stefan-Boltzmann constant,
and $D$ is the distance from the source to the observer. 
The model atmospheres were constructed following 
standard temperature-correction procedures as outlined in Mihalas (1978)
in order to satisfy the hydrostatic and radiative equilibrium constraints.
The 15 most abundant 
elements from H~through Ni~are included with bound-free opacities 
computed from the photoionization cross sections of Verner \etal\ (1993,1995). 
Line blanketing is included using 
$\sim$1200 of the strongest spectral lines
from the CHIANTI (v.3.0) atomic database (Dere \etal\ 1997).

A series of LTE model atmosphere spectra were constructed in this fashion
spanning the range of parameters $10^5 \le T_{eff} \le 1.3 \times 10^6$ and  
$7.5 \le$~log($g$)~$\le 9.5$, as appropriate for WD atmospheres, 
with the scale factor $K$ used as the 
model normalization in XSPEC. 
Tables of these models were used in conjunction with a photoelectric
absorption component in fitting to the observed spectra.  
A separate table was 
constructed for each of two values of photospheric metal abundances, 
$Z/Z_{\odot}=1$ and $Z/Z_{\odot}=0.01$, relative to the
solar composition given by Anders \& Grevesse (1989). The lower metal
abundance is roughly compatible with the observed M81 metallicities 
($\sim$0.03) recently reported by Kong \etal\ (2000).
In addition, a He-rich model consisting of $Z/Z_{\odot}=0.03$ metals 
and 99\% He by weight was also constructed. 

For reference, computed hot WD LTE model atmosphere 
spectra at several temperatures are
shown in Figure~\ref{f:lte_models}. 
Many of the edges seen in these spectra will prove key diagnostics of
the properties of the candidate \ssss.

\begin{center}
\includegraphics[angle=90,width=\columnwidth]{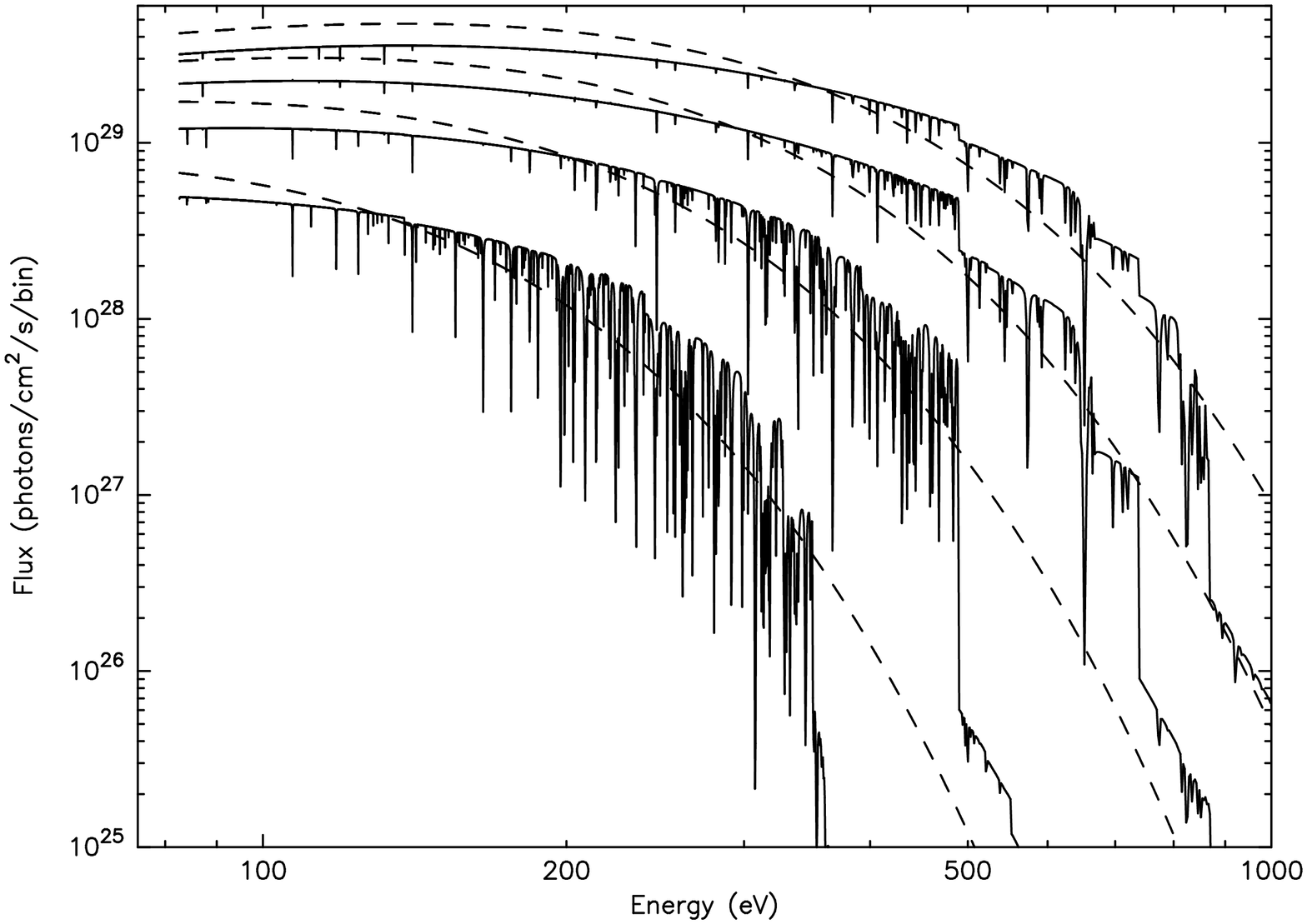}
\vspace{10pt}
\figcaption{Hot white dwarf LTE atmosphere model spectra ({\em solid}) and 
corresponding blackbody spectra ({\em dash}) at four effective temperatures
and a solar abundance composition. From top to bottom,
the temperatures are $T_{eff}/10^5\rm{K} = 10$, 8, 6, and 4.
The surface gravity is log($g$)=9 
except for the hottest model where log($g$)=9.5 to avoid exceeding
the Eddington limit luminosity. Prominent bound-free edges are apparent at
490 eV (C VI), 670 eV (N VII), 740 eV (O VII), and 870 eV (O VIII).
\label{f:lte_models}}
\end{center}

\begin{table*}
\begin{center}
\small{
{\sc Table 2} \\
{\sc Model Fits for Source N1} \vspace{6pt} \\ 
\begin{tabular}{cccccccc} 
%
\hline \hline
Model$^a$                  & blackbody             & blackbody             &  blackbody            & LTE                 & LTE               & LTE $+$ ray   & diskbb \\
                           &                       & $+$ edge              &  w/ zvarabs$^b$       & $Z=Z_{\odot}$       & $Z=0.01Z_{\odot}$ & $Z=Z_{\odot}$ &        \\ \hline 
$\chi_{\nu}^2$/dof         & 1.40/47               & 0.98/45               & 1.05/44               & 2.75/46             & 3.55/47          & 1.24/44        & 1.61/47\\
$N_H/10^{20}$~cm$^{-2}$    & $5.7^{+0.7}_{-0.6}$   & $4.2^{+1.0}_{-1.3}$   & $6.3^{+0.5}_{-0.5}$   & $4.2^{+0.6}_{-0.4}$ & $4.0^{+0.6}_{-0.4}$ & $4.4^{+0.5}_{-0.6}$ & $7.4^{+0.8}_{-0.8}$ \\
$T_{eff}$~eV               & $86^{+2}_{-3}$        & $81^{+3}_{-4}$        & $78^{+3}_{-3}$        & $69^{+1}_{-1}$      & $64^{+2}_{-1}$   &  $66^{+1}_{-1}$& $100^{+4}_{-3}$\\
$E_{edge}$~keV, $\tau$     & ---                   & $0.26^{+0.02}_{-0.02}$, 1.6 & ---             & ---                 & ---              & --- & --- \\
 C,O enhancement           & ---                   & ---                   & $6.6^{+1.7}_{-1.7}$, $1.5^{0.9}_{2.1}$              & ---                 & ---              & --- & --- \\
log($g$)~cm$^2$~s$^{-1}$   & ---                   & ---                   & ---                   & $8.7^{+0.04}_{-0.04}$ & $8.8^{+0.05}_{-0.03}$ & $8.5^{+0.03}_{-0.1}$& --- \\
thermal $kT$, $f(L_{X})^c$ & ---                   & ---                   & ---                   & ---                   & --- &$0.35^{+0.12}_{0.07}$, 0.18 & --- \\
$L_{obs}/10^{38}$~\ergl\   & $3.0^{+0.02}_{-0.02}$ & $3.0^{+0.04}_{-0.04}$ & $2.9^{+0.02}_{-0.03}$ & $2.9^{+0.02}_{-0.02}$ & $2.9^{+0.04}_{-0.04}$ & $3.0^{+0.01}_{-0.01}$ & $3.0^{+0.03}_{-0.03}$\\
$L_{X}/10^{38}$~\ergl\     & $8.7^{+1.6}_{-1.2}$  & $7.3^{+2.9}_{-1.7}$   & $8.8^{+1.4}_{-4.7}$  & $6.3^{+1.4}_{-1.4}$   & $6.2^{+1.6}_{-1.5}$   & $6.7^{+0.6}_{-0.6}$   & $13.4^{+3.0}_{-2.9}$\\
$L_{bol}/10^{38}$~\ergl\   & $12.0^{+2.0}_{-2.1}$ & $15.0^{+5.2}_{-1.1}$  & $23.0^{+11.2}_{-8.0}$ & $9.0^{+2.0}_{-1.1}$  & $8.0^{+1.3}_{-1.1}$  & $9.0^{+2.4}_{-2.1}$  & $11.0^{+1.4}_{-1.0}$/cos$i$ \\
\hline 
\multicolumn{8}{l}{$^a$All models include solar abundance absorption column} \\
\multicolumn{8}{l}{$^b$variable-abundance absorption column with best-fit redshift $z=0.04$} \\
\multicolumn{8}{l}{$^c$Raymond-Smith model temperature (keV) and fraction of $L_{X}$ due to this component} \\
\end{tabular}
} 
\end{center}
\end{table*}

\subsection{Source N1}

The spectral fit results for the brightest \sss\ candidate, source N1,
are listed in Table~2. 
This and subsequent tables for the other bright sources 
include one column for each model. Rows list the $\chi^2$ fit statistic,
the fitted model parameters
$N_H$ (hydrogen column density), $T_{eff}$ (blackbody or effective temperature),
log($g$) (surface gravity), $E_{edge}$ (absorption edge energy) 
and the accompanying $\tau$ (edge optical depth). All models include 
photoelectric absorption from a solar abundance column.
In the case of model {\em zvarabs}, the abundance of C and O in the 
absorbing column are allowed to vary. The table lists the enhancement
of these elements relative to the solar values.
Also listed are the observed luminosity, $L_{obs}$, in the 0.2--2.0~keV band,
the unabsorbed luminosity, $L_{X}$, in the same band and the 
bolometric luminosity, $L_{bol}$, as scaled from the 
model normalizations assuming a distance of 3.6 Mpc to M81.

The observed spectrum, best-fit absorbed blackbody model with added 
absorption edge at 0.26~keV (second model of Table~2), and fit
residuals for source N1 are shown in Figure~\ref{f:sn1_spectralfit}.
A simple blackbody model is statistically superior to either of the LTE WD
atmosphere models because of the lack of absorption edges in the
observed spectrum. 
However, there is a feature in the blackbody fit residuals near 
0.3 keV.  This feature can
be modelled as an edge at $0.26\pm0.01$~keV (second column of Table 2) 
which is very close to the neutral C K-edge
at 0.277~keV.  Since there is no reason
to expect the galactic absorbing column to have exactly solar
abundances, we also tried a variable abundance model.  If a small
"redshift" (z=0.04) is applied, this model also provided an excellent fit
to the data as shown by the third column in Table 2.  It is worth
noting that the ACIS energy scale is uncertain at the lowest energies
and thus the feature is likely a Carbon edge and the "redshift" is
due to this uncertainty.  As the ACIS response
is also uncertain at these energies we cannot determine whether
the excess Carbon is intrinsic to the source in M81, lies within our galaxy,
or is a detector effect.


\begin{center}
\includegraphics[angle=90,width=\columnwidth]{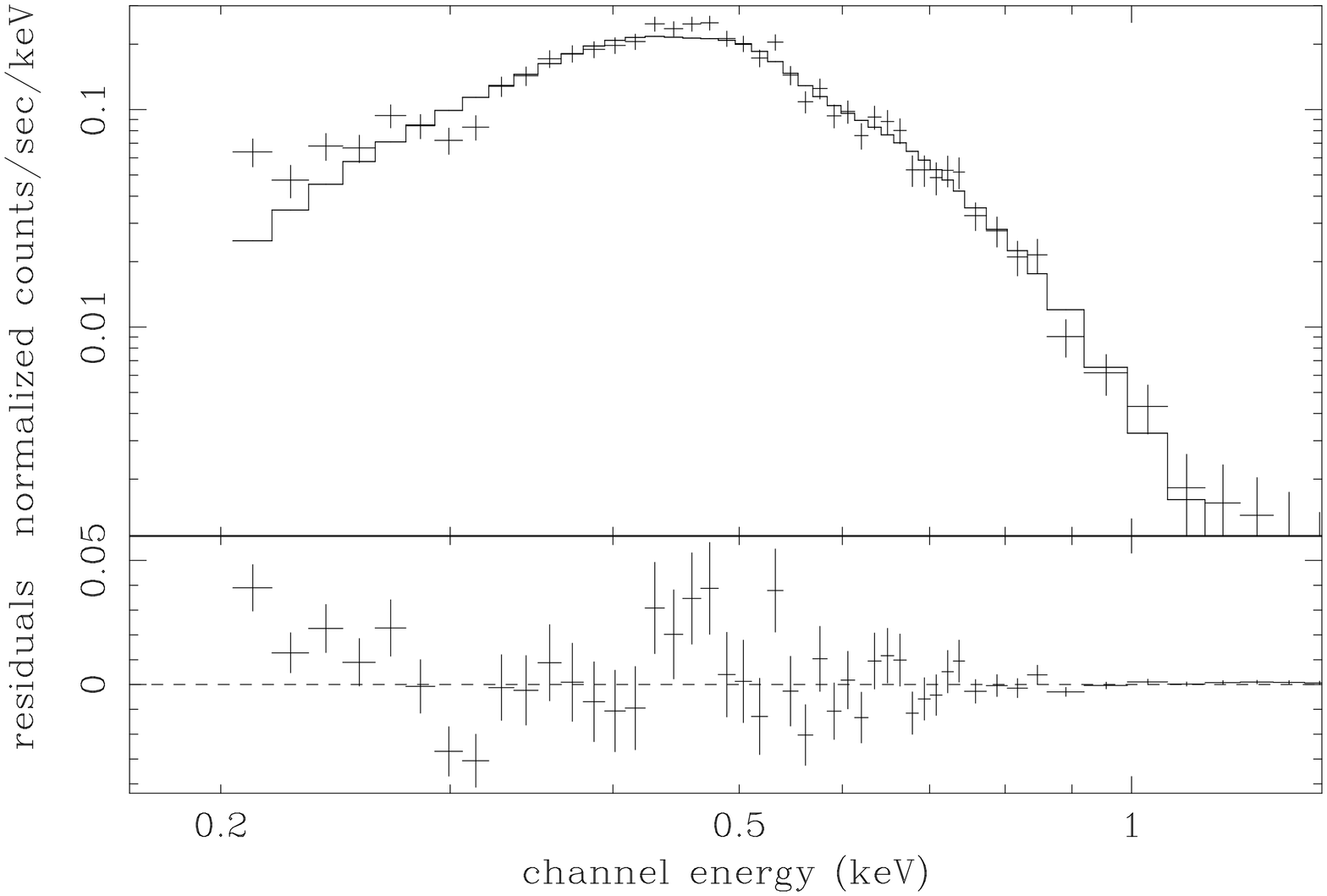}
\vspace{20pt}
\figcaption{{\em Upper panel:} Observed spectrum of source N1 (symbols) and 
best-fit blackbody model with solar-abundance photoelectric absorption 
and added absorption edge at 0.26~keV (solid line). 
{\em Lower panel:} Fit residuals.
\label{f:sn1_spectralfit}}
\end{center}

LTE models with an additional Raymond-Smith thermal
emission component are an improvement over the single LTE component
models because this added thermal component tends to fill the $\sim$0.7--1.0~keV
region of the model spectrum underestimated due to the presence 
of absorption edges in the LTE component.
While the physical basis for such a thermal component has precedence in
modeling of, {\em e.g.}, the recurrent nova U~Sco (Kahabka \etal\ 1999)
and the symbiotic nova SMC~3 (Jordan \etal\ 1996),
and there is reason to believe source N1 may be near a state of dynamical
outflow (see discussion in \S\ref{s:discussion}), the fact that a simple 
blackbody model provides an equally-acceptable fit allows one to conclude 
that the thermal component is not necessary.
It should be noted that $\sim$2\% of the incident flux 
at 0.5~keV is piled up in the spectrum of source N1
resulting in detected events near 1~keV. 
The observed flux at 1.0--1.1~keV is
3.5\% of that in the 0.5--0.55~keV band. Thus some 60\% of the observed flux
at 1.0--1.1~keV is due to pileup.


He-rich LTE model atmospheres were also fitted to this spectrum
but no statistically-acceptable
fits were obtained and they are omitted from Table~2 for brevity. 
The larger He opacity near the HeII edge in these models leads to
a slightly lower temperature and a slightly steeper temperature gradient
in spectrum-forming (0.1$\LA\tau_R\LA$1, where $\tau_R$ is the Rosseland
mean opacity) regions of the atmosphere. 
The emergent
flux therefore decreases near the HeII edge (due to lower temperatures) 
and increases slightly at the peak of the spectral 
energy distribution (due to the steeper gradient). The overall effect
in the \cha\ observable energy band is to produce a steeper spectrum  
than the H-rich LTE models.
The spectrum is further steepened by the bound-free edges 
from intermediate-mass metals resulting in a poorer model of the data.

The blackbody shape of the observed spectrum of source N1 is
reminiscent of that expected from a geometrically thin optically thick 
accretion disk surrounding a black hole
(Shakura \& Sunyaev 1973). 
For completeness, therefore, an accretion disk emission model was also 
fitted to the observations. This model accumulates the
spectrum from an optically thick accretion disk observed at 
inclination angle $i$ from a superposition of blackbody spectra from 
concentric annular elements representing the radial dependence of the 
disk temperature profile (Mitsuda \etal\ 1984, Makishima \etal\ 1986).
The model parameters (model {\em diskbb}, Table 2) are the
normalization, proportional to the innermost disk radius, $R_{in}$, 
for a known inclination angle and distance to the source, 
and the temperature at $R_{in}$.
Assuming $R_{in}$ corresponds to the last stable Keplerian orbit 
(Makishima \etal\ 2000) implies a mass for the central object. 
The best-fit normalization for source N1 corresponds to a black hole mass 
$M_{BH}$$\sim$1200/(cos$i)^{-1/2}$ assuming the last stable orbit is at
3~Schwarzschild radii.
The best-fit temperature at $R_{in}$, $T_{in}$$\sim$100~eV, implies a 
bolometric luminosity $L_{bol}$$\sim$$10^{39}$~\ergl.
Not surprisingly, the color temperature $T_{in}$
is slightly higher than that obtained from the blackbody models.
The shape of the blackbody disk model is a power law of photon index $-2/3$
at low photon energies steepening to the Wien portion of a blackbody 
of temperature $T_{bb}=0.7T_{in}$ at higher energies 
(Makishima \etal\ 1986). Thus the combination of an innermost
disk radius temperature slightly higher than the blackbody model 
temperature and
a higher absorption column density is just what is needed to make the 
disk blackbody model mimic a standard blackbody spectrum. 
This is especially true in view of the fact that only the steep 
Wien portion of the spectrum lies within the \cha\ energy band 
for low values of $T_{in}$ and the power law portion of the 
blackbody disk model spectrum lies entirely below the low-energy
instrumental cutoff. The lack of detectable 
emission at longer wavelengths at the position of source N1 
(\S\ref{s:environments}) and the lack of an observable hard power-law tail 
argues against the blackbody disk model.

The luminosity of source N1 is exceptionally high. 
The observed 0.2--2.0~keV luminosity, 
$L_{obs}$$\sim$$3\times 10^{38}$~\ergl, is slightly
above the Eddington limit for a 1.4 \msun\ H-accreting star. 
The unabsorbed luminosity on the same energy range and the inferred bolometric
luminosities are much higher though with high uncertainties.
In contrast, $L_{obs}$ is tightly constrained and its value does not 
vary among the models (Table~2).
Another source of uncertainty,
independent of model fit uncertainties, is the gain behavior in the
BI device at low photon energies and at the -120$^{\circ}$C
focal plane temperature in use at the time of observation.
Very nearly 1/3 of the observed flux from source N1
falls in the 0.2--0.4~keV energy band. 
A 20\% uncertainty in the flux below 0.4~keV 
corresponds, therefore, to an additional 7\% uncertainty in $L_{obs}$.
Conservatively, then, the estimated bolometric luminosity of source N1 is 
at least several times $10^{38}$ \ergl\ and, depending on
the model, may be as high as a few $10^{39}$ \ergl.
This result places severe constraints on
physical models for this source (\S\ref{s:discussion}). 

It is notable that the large range of $L_X$ and $L_{bol}$
is not obviously reflected in a large range
of the model fit parameters $N_H$ and $T_{eff}$. 
From Table~2, the range of these parameters among the
various models (excluding {\em diskbb}) are $N_H = 4.9\pm2.0\times 10^{20}$~cm$^{-2}$ and
$T_{eff}= 75\pm11$~eV. This substantiates the previous conjecture 
that the inferred luminosities are very sensitive to small variations
in the model parameters.
However, the differences in shape of the blackbody
and LTE model atmosphere spectra do 
not give systematically large differences in the 
derived temperatures, column densities, and luminosities
as is often reported for \ssss. 

The hot WD LTE model parameter log($g$) is directly related to the
mass of the assumed white dwarf. The resulting white dwarf
mass is estimated to be 0.9 to 1.1 \msun\ for source N1. 

\subsection{Source N2}

The spectral fit results for source N2
are listed in Table~3. 
The observed spectrum, best-fit LTE atmosphere model, and fit
residuals are shown in Figure~\ref{f:sn2_spectralfit}.

\begin{center}
\small{
{\sc Table 3} \\
{\sc Blackbody Model Fits for Source N2} \vspace{6pt} \\ 
\begin{tabular}{ccc} 
\hline \hline
Model$^a$                  & blackbody             & blackbody with edge   \\ \hline
$\chi_{\nu}^2$/dof         & 1.87/16               & 0.93/14               \\
$N_H/10^{20}$~cm$^{-2}$    & $19.8^{+8.0}_{-7.5}$ & $9.3^{+5.2}_{-5.7}$  \\
$T_{eff}$~eV               & $98^{+15}_{-11}$      & $171^{+60}_{-29}$      \\
$E_{edge}$~keV, $\tau$     & ---                   & $0.85^{+0.01}_{-0.03}$, 3.8   \\
$L_{obs}/10^{38}$~\ergl\   & $0.4^{+0.02}_{-0.02}$ & $0.5^{+0.04}_{-0.04}$ \\
$L_{X}/10^{38}$~\ergl\     & $3.2^{+5.9}_{-2.1}$   & $0.9^{+1.4}_{-0.3}$   \\
$L_{bol}/10^{38}$~\ergl\   & $4.0^{+4.9}_{-0.5}$   & $1.3^{+0.7}_{-0.3}$   \\
\hline 
\vspace{6pt} \\
%
\multicolumn{3}{c}{{\sc LTE Atmosphere Model Fits for Source N2}} \vspace{6pt} \\
%
\hline \hline
Model$^a$                  & $Z=Z_{\odot}$        & $Z=0.01Z_{\odot}$     \\ \hline
$\chi_{\nu}^2$/dof         & 1.38/15              & 1.00/15               \\
$N_H/10^{20}$~cm$^{-2}$    & $25.0^{+7.1}_{-8.4}$ & $15.8^{+3.8}_{-4.2}$  \\
$T_{eff}$~eV               & $69^{+4}_{-1}$       & $67^{+2}_{-1}$        \\
log($g$)~cm$^2$~s$^{-1}$   & $8.49^{+0.03}_{-0.03}$ & $8.24^{+0.40}_{-0.08}$ \\
$L_{obs}/10^{38}$~\ergl\   & $0.3^{+0.01}_{-0.01}$ & $0.4^{+0.03}_{-0.03}$  \\
$L_{X}/10^{38}$~\ergl\     & $4.5^{+1.4}_{-3.7}$  & $2.0^{+1.5}_{-0.8}$   \\
$L_{bol}/10^{38}$~\ergl\   & $6.4^{+3.9}_{-1.3}$  & $5.3^{+4.8}_{-2.1}$   \\
\hline
\multicolumn{3}{l}{$^a$All models include solar abundance absorption column} \\
\label{t:sn2}
\end{tabular}
} 
\end{center}

\begin{center}
\includegraphics[angle=90,width=\columnwidth]{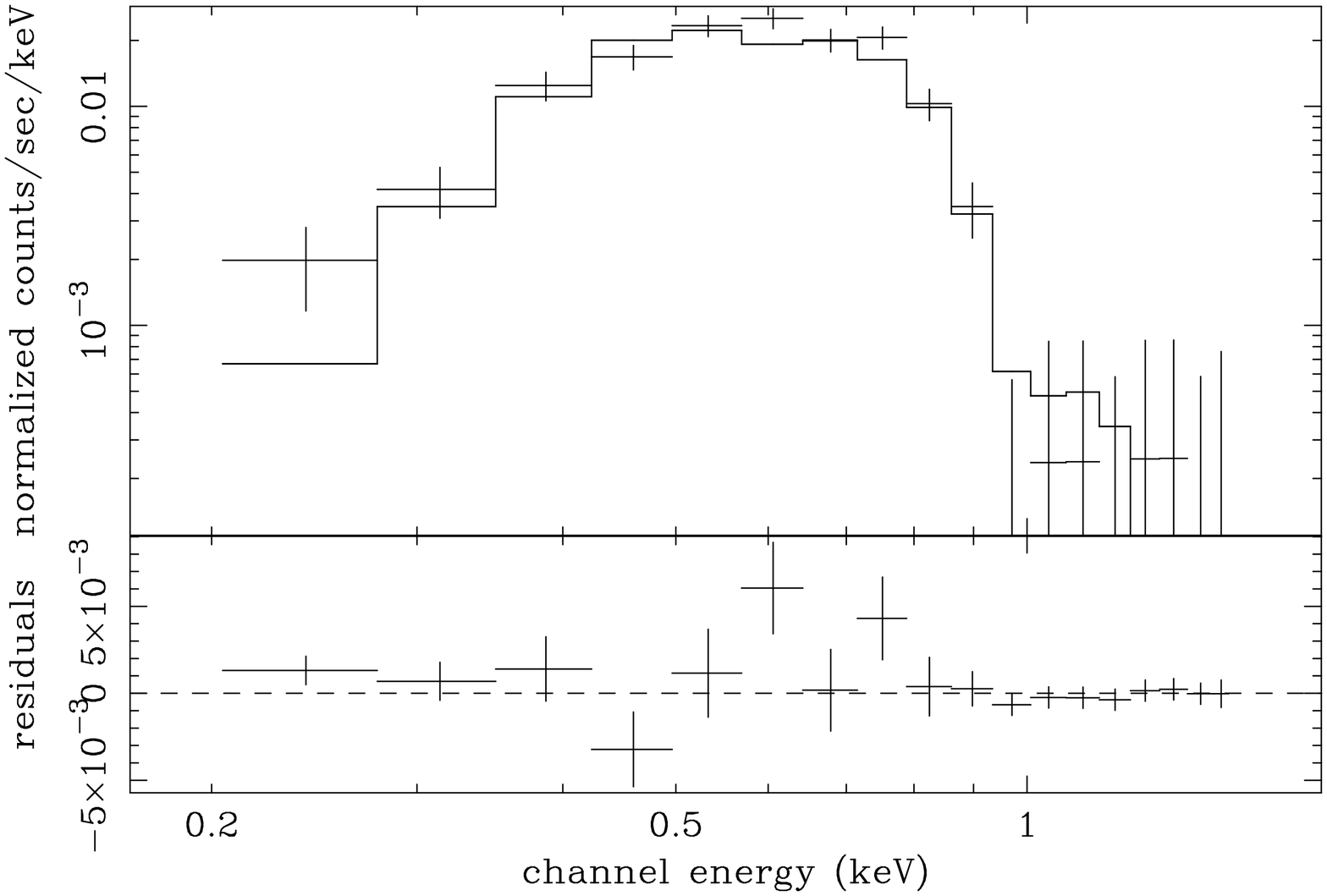}
\vspace{20pt}
\figcaption{{\em Upper panel:} Observed spectrum of source N2 (symbols), 
best-fit LTE WD model atmosphere spectrum 
with 1\% solar abundance metals 
and solar-abundance photoelectric absorption 
(solid line). {\em Lower panel:} Fit residuals.
\label{f:sn2_spectralfit}}
\end{center}

In contrast to source N1, a blackbody model 
is clearly a poor fit to the data. 
A blackbody with an added absorption edge at 
$0.85^{+0.01}_{-0.03}$~keV gives a better fit suggesting  
the presence of highly-ionized O in a very hot atmosphere.
The LTE models provide significant improvements over the 
blackbody model. The results are 
not very sensitive to the assumed metal abundances though
the 1\% solar metal abundance model provides an improved
fit. At the derived temperatures, $T_{eff} \sim 67$--69~eV,
the hot WD LTE models produce numerous absorption edges. 
When convolved with the instrument response, the most prominent
edges are those from highly ionized O at 740 eV (O VII) and 870 eV (O VIII). 
In the low-abundance model, the edges are
not as deep and an additional Ne IX edge (1196~eV) appears in the model
but is statistically insignificant in the fit. 

The observed 0.2--2.0~keV luminosity derived from the hot WD 
LTE model atmosphere spectrum is 
much less than the Eddington limit for a hydrogen-accreting WD. 
The inferred range of bolometric luminosities
exceed this limit by a factor of 2 to 3.
The white dwarf mass inferred from the LTE models is 0.7 to 0.9 \msun.

\subsection{Source N3} \label{s:sn3}

The spectral fit results for source N3
are listed in Table~4. 
The observed spectrum, best-fit hot WD LTE model atmosphere spectrum, 
and fit residuals are shown in Figure~\ref{f:sn3_spectralfit}.

\begin{center}
\small{
{\sc Table 4} \\
{\sc Model Fits for Source N3} \vspace{6pt} \\ 
\begin{tabular}{ccccc} 
\hline \hline
Model$^a$                  & blackbody            & $Z=Z_{\odot}$        & $Z=0.01Z_{\odot}$     \\ \hline
$\chi_{\nu}^2$/dof         & 0.94/4               & 0.90/3               & 1.4/3                \\
$N_H/10^{20}$~cm$^{-2}$    & $10.7^{+12.6}_{-7.1}$ & $4.7^{+5.3}_{-3.2}$ & $8.7^{+5.4}_{-4.3}$  \\
$T_{eff}$~eV               & $52^{+13}_{-5}$      & $52^{+1}_{-1}$       & $45^{+9}_{-6}$        \\
log($g$)~cm$^2$~s$^{-1}$   & ---                  & $7.9^{+0.9}_{-0.1}$  & $9.2^{+0.2}_{-1.6}$ \\
$L_{obs}/10^{38}$~\ergl\   & $0.1^{+0.03}_{-0.03}$ & $0.1^{+0.02}_{-0.02}$  & $0.1^{+0.03}_{-0.02}$   \\
$L_{X}/10^{38}$~\ergl\     & $2.0^{+4.5}_{-1.3}$  & $0.4^{+0.3}_{-0.2}$  & $1.0^{+1.9}_{-0.9}$   \\
$L_{bol}/10^{38}$~\ergl\   & $4.6^{+15.3}_{-3.4}$ & $1.2^{+1.3}_{-0.7}$  & $1.8^{+8.0}_{-1.2}$  \\
\hline
\multicolumn{5}{l}{$^a$All models include solar abundance absorption column} \\
\label{t:sn3}
\end{tabular}
} 
\end{center}

\begin{center}
\includegraphics[angle=90,width=\columnwidth]{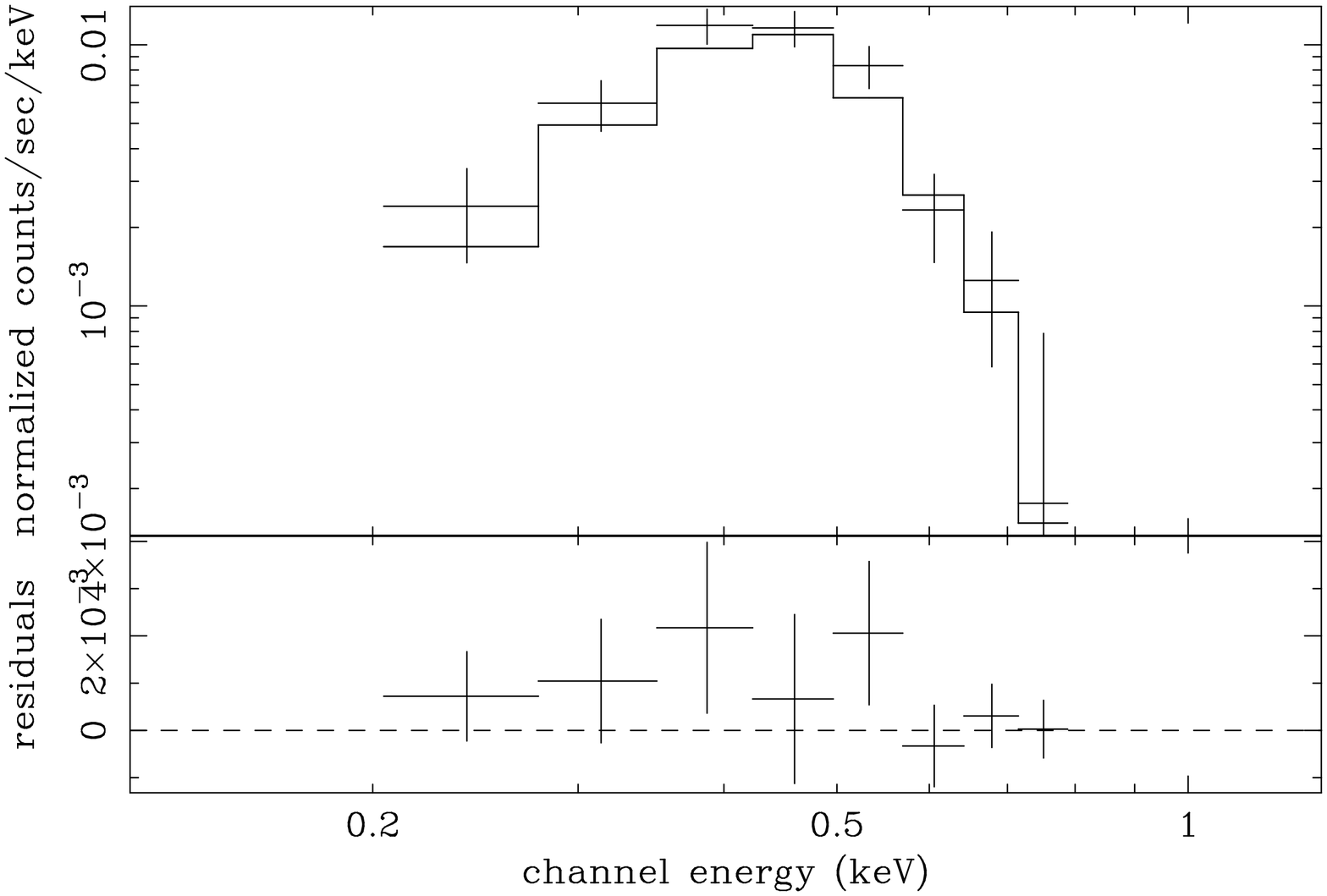}
\vspace{20pt}
\figcaption{{\em Upper panel:} Observed spectrum of source N3 (symbols), 
best-fit WD LTE atmosphere model spectrum with solar abundance metals 
and solar-abundance photoelectric absorption 
(solid line). Spectrum and model have been rebinned to show structure. 
{\em Lower panel:} Fit residuals.
\label{f:sn3_spectralfit}}
\end{center}

Unlike the stronger sources, the spectrum of source N3 is fit
equally well by all three models 
and there is no spectral feature
to meaningfully discriminate
between models. In particular, the lack of flux above $\sim$0.7~keV
means that the dominant absorption edges present in the LTE atmosphere
models are the relatively weak edges at C VI (490~eV) and N VII (670~eV).
In this case, a slightly more absorbed blackbody model relative to the
LTE models adequately 
mimics the observed steep decline above $\sim$0.5~keV.

The observed 0.2--2.0~keV luminosity, $L_{obs}$, is
less than the Eddington limit for a hydrogen-accreting WD and
$L_{bol}$ is comparable to the Eddington limit though, again, it is 
highly uncertain.
The white dwarf mass inferred from the WD atmosphere LTE models 
is $\sim$0.5 \msun\ for the solar
abundance model and is effectively unconstrained 
by the 1\% solar metallicity model.

\subsection{Other Sources} \label{s:remaining_spectra}

Statistically constrained model fits to the remaining sources 
could not be achieved. The observed count rates and similarities to
the spectra of the brighter sources allow a crude extrapolation 
of the derived luminosities of
the brighter sources to the remaining sources by the following reasoning: 
The ratio $L_{bol}/L_{obs}\sim 5$ for source N1 and $\sim 10$--20
for sources N2 and N3. The latter two sources are associated with
spiral arms. They have higher absorbing columns than source N1
which accounts for the higher $L_{bol}/L_{obs}$ ratio.
An estimate of $L_{bol}$ can be made by assuming the ratio
$L_{bol}/L_{obs} = 15$ for sources on the spiral arms and
$=5$ otherwise and noting that $L_{obs}$ is proportional to the 
observed count rate (Table~1). 
Applying this approximation to the remaining sources using the 
data from Table~1 gives 
the following values of $L_{bol}$ in units of $10^{37}$ \ergl: 
2.2 (N4), 1.9 (N5), 1.6 (N6), 1.3 (N7), 3.6 (N8), and 2.4 (N9).
The uncertainties in these estimates are at least a factor of three
since the actual absorbing columns are unknown. This uncertainty
is, alas, comparable to some of the uncertainties in $L_{bol}$
obtained for the brightest sources.

Alternatively, luminosities can be estimated from the count rate
and assuming a (blackbody) temperature and a column density.
The weakest \sss\ detected has an 
observed luminosity of $2\times 10^{36}$~\ergl 
assuming an 80~eV blackbody 
spectrum and a column of $N_H=8\times 10^{20}$~cm$^{-2}$. 
This corresponds to $L_{bol} \sim 1.3\times 10^{37}$~\ergl.
$L_{bol}$ increases to $5.2\times 10^{37}$~\ergl\ if the assumed
temperature is lowered to 50~eV.

Temperatures cannot be so easily estimated but they are most 
probably of order 40-50 eV as higher temperature sources are 
expected to be rare based on theoretical models of \ssss\ 
(see \S\ref{s:discussion})
and cooler sources are difficult to detect with \cha-ACIS.
Even an intrinsically bright ($L_{bol}=10^{38}$~\ergl) 
blackbody source with
$T_{eff}=25$~eV and moderate absorbing column (2 times Galactic) 
produces only 24 counts
in the BI device and would be undetectable in the FI chips. 

\section{X-ray Timing Analysis} \label{s:lc_analysis}

\subsection{Short-Term Variability}

The 50~ks \cha\ light curves of the brightest 3 \sss\ candidates 
are presented in Figure~\ref{f:cha_lc}. Detected events have been binned on 1000~sec intervals.

\begin{center}
\includegraphics[angle=90,width=\columnwidth]{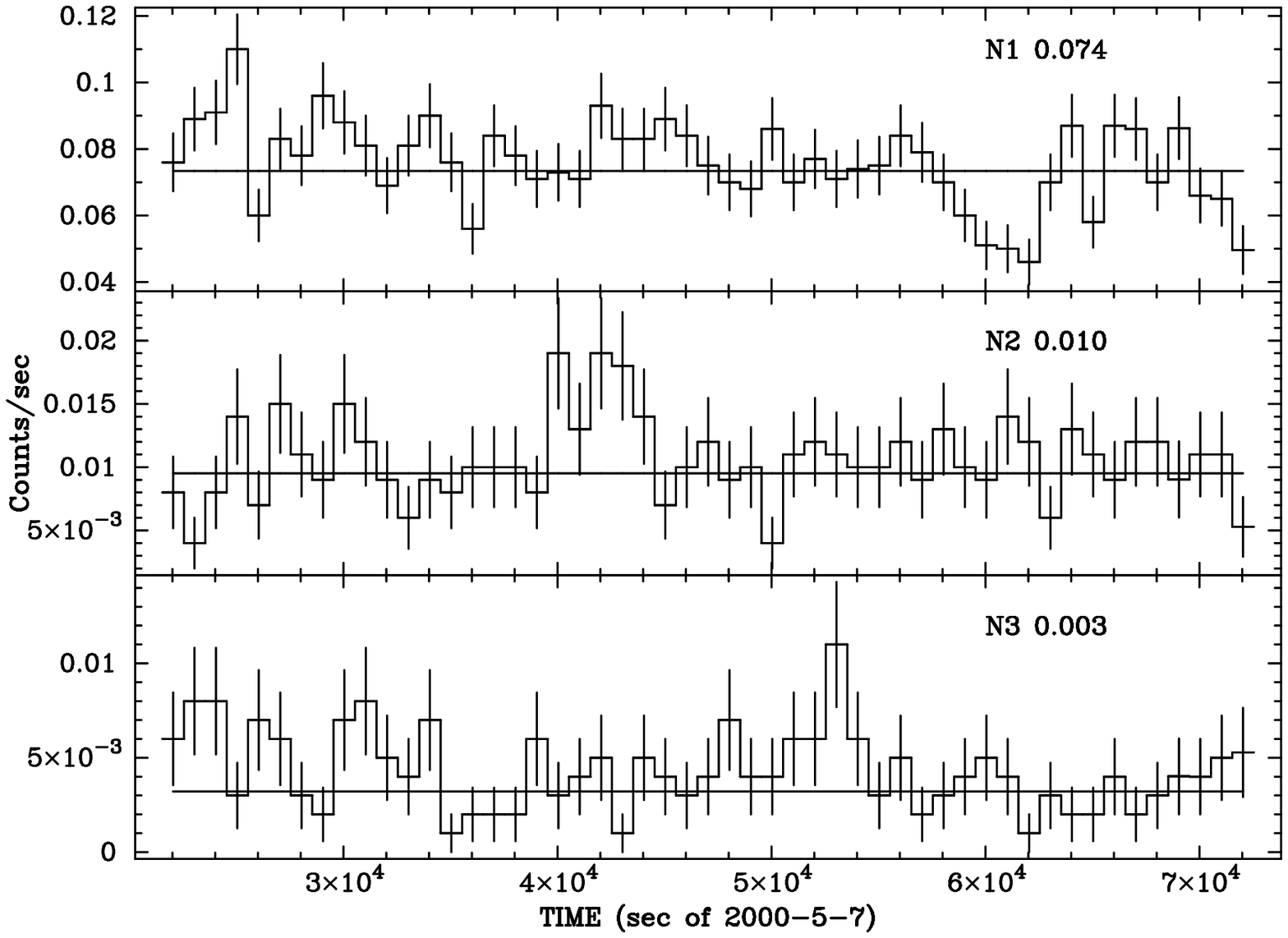}
\vspace{20pt}
\figcaption{\cha\ light curve of the brightest \sss\ candidates binned
on 1000~sec intervals. All standard-grade events in the 0.2--2.0~keV bandpass, 
within valid good-time intervals, and within the same spatial regions 
as used for spectral analysis, are included. Error bars represent 
1$\sigma$ statistical uncertainties. Background contributions are
negligible: Typical background count rates are 
10$^{-6}$ counts~s$^{-1}$~pixel$^{-1}$. Labels denote source number and 
average count rate obtained by fitting a constant to the data
(shown as horizontal lines through the data).  \label{f:cha_lc}}
\end{center}

The observation was of sufficient 
duration that the Kolomogorov-Smirnov statistic can be evaluated for the 
brightest sources to test the hypothesis that the sources are constant.
The test shows sources N1 and N2 are variable at high confidence but the
remaining sources contain insufficient data to be conclusive. Power spectra
were also generated for the brightest sources to search for pulsations or 
other periodic behavior. Only source N1 had sufficient signal to obtain 
significant results. No periodicity was detected.

An additional 2.4~ks ACIS-S imaging observation of M81 obtained 
on 2000 Mar 21, 47 days previous to the 
50~ks observation, was acquired through the \cha\ X-ray Center data archive. 
Source N1 is the only \sss\ candidate with a sufficient number of counts
in this short exposure for analysis.
The light curve of source N1 during this observation, binned on 300~sec
intervals, is shown in Figure~\ref{f:2ks_lc}. Also shown are the 
corresponding light curves in the energy bands 0.2--0.5 and 0.5--1.0~keV
and the time dependence of the background during the observation.
The source brightness declines from a maximum at the beginning of 
the observation to a statistically insignificant level 
within $\sim$900~s and remains at this low level 
for the remainder of the observation. No other (bright) source displayed 
similar changes in X-ray flux during the 2.4~ks observation.

\begin{center}
\includegraphics[angle=90,width=\columnwidth]{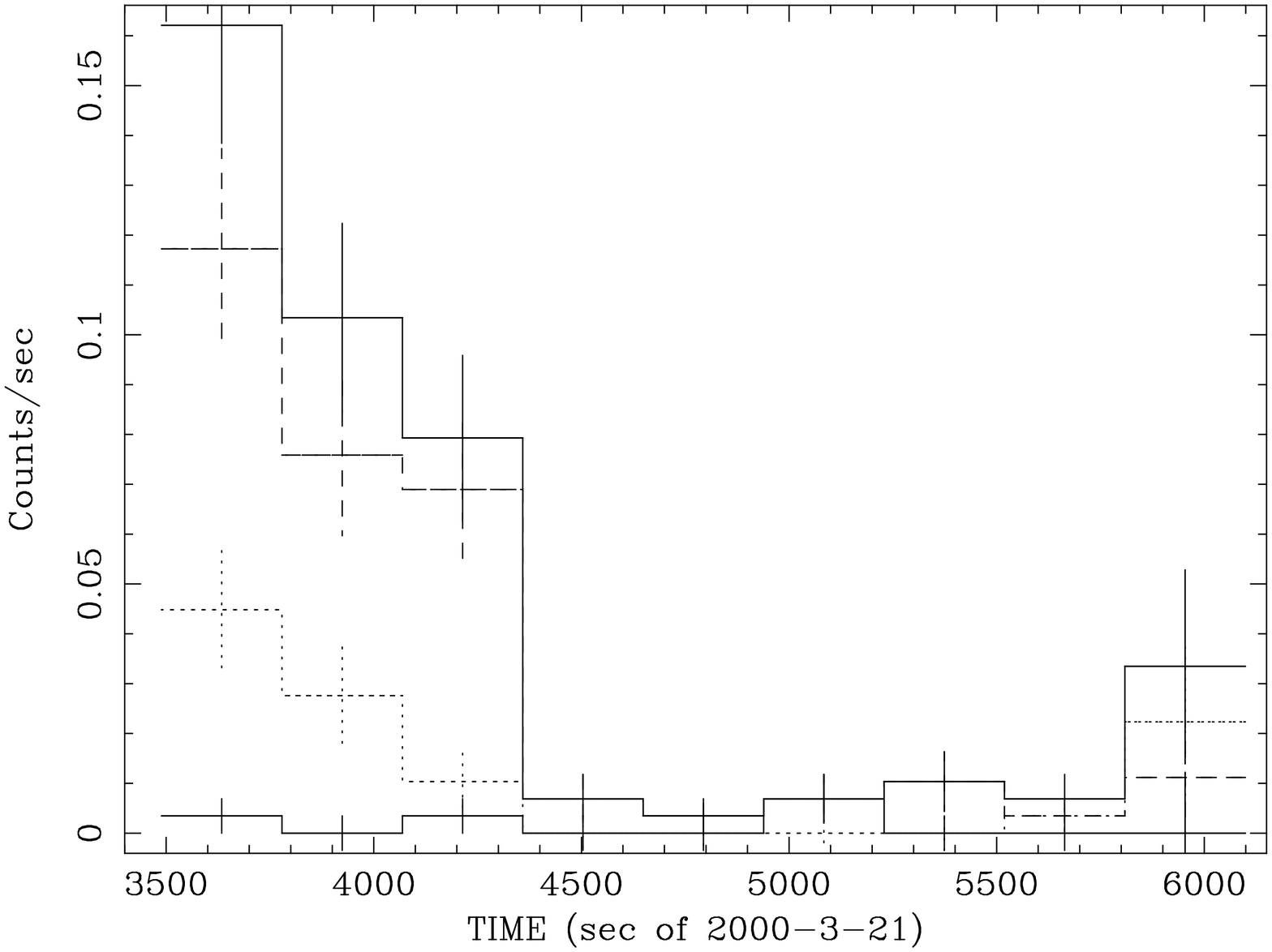}
\vspace{20pt}
\figcaption{\cha\ 2ks light curve of source N1 binned on 300~sec intervals. 
The uppermost curve includes all standard grade
events in the 0.2--2.0~keV band,
the next lower curve represents the 0.5--1.0~keV band ({\em dashed})
followed by the 0.2--0.5~keV band ({\em dotted}).
The lower curve represents the nearby background 
scaled by the ratio of the source to the background area. Error bars represent 
1$\sigma$ statistical uncertainties.
\label{f:2ks_lc}}
\end{center}

The dramatic drop in flux suggests an eclipse event.
The observation interval is too short to
establish a brightening of the source signifying the end of the
eclipse 
(the observation ended 100~sec into the last interval and this
interval contains 4 counts, one of which is probably a background
photon based on its PI value).
No similar feature is detected in the longer \cha\
observation and no similar feature can be seen in the 
relatively poor signal-to-noise 
\ros\ data discussed below.

Assuming the drop in flux signals the onset of an eclipse of 
an $\sim$1 \msun\ compact object,
that Roche-lobe overflow from the companion is occuring,
that the eclipse duration exceeds 1600~s,
and that no eclipse occurs during the 50~ks exposure,
weak constraints on the orbital elements of any binary system can be 
derived. 
These constraints
are easily met by massive companions ($\GA$3 \msun) that have 
evolved off the main sequence. Systems in this mass range have periods
greater than $\sim$1.6 days and eclipse fractions of order 15\%. Missing
an eclipse is then quite likely though witnessing the onset of an 
eclipse in a short exposure would be fortuitous. 
Furthermore, such companion
stars should be visible in the \hst\ images (\S\ref{s:environments}).  
A more likely scenario for \ssss\ is a slightly-evolved 
main sequence donor in the mass 
range 1.3--2.5~\msun\ (van~den~Heuvel \etal\ 1992) or a ``helium Algol''
with a few-solar-mass companion at the
onset of He accretion (Iben \& Tutukov 1994, Yungelson \etal\ 1996).
Orbital periods range from 10~hr to 30~hr for the main sequence donor scenario
and $<$20~hr for the He-accreting systems (Yungelson \etal\ 1996).
There is at best a one-in-three chance that 
no eclipse is observed during a 50~ks (13.9~hr)
observation of systems with a 30~hr period and the probability decreases
for the shorter-period systems. 
Thus, though an eclipse
cannot be excluded by the X-ray observations it is not very likely.

Assuming the source is exiting an eclipse phase during the last $\sim$300~s
of the short exposure, then the eclipse lasts only of order 1800~s.  
The orbital period must 
then be of order a few hours only since $\sim$15\% of the period is spent in 
eclipse. A period this short could not be missed in the 50~ks observation.

Alternatively,
nuclear burning on the surface of a WD may be occuring through weak
shell flashes that either cause expansion which quenches 
burning or expels material which obscures the underlying atmosphere
as it cools with more violent flashes followed by a larger flux decline. 
This may explain the Figure~\ref{f:2ks_lc} light curve where
the X-ray flux is roughly a factor-of-two brighter during the onset 
of the observation than the average value observed during the 
50~ks observation (Figure~\ref{f:cha_lc}).
Fitting an exponential to the first 1000~s of this light curve 
gives an e-folding timescale of $\sim 350$~s. This is comparable to the sound
crossing time over the surface of a WD for reasonable estimates of the 
sound speed in the burning layer.
A change in accretion rate may also trigger large excursions in the 
X-ray brightness. Southwell \etal\ (1996) have suggested this mechanism may 
explain the factor of $\sim$20 rise in X-ray flux observed
(Schaeidt, Hasinger, \& Tr\"{u}mper 1993) in the \sss\ RXJ~0513.9-6951.
However, the timescale for contraction of the photosphere leading to the 
required change in $T_{eff}$ is of order days (Livio 1992). RXJ~0513.9-6951
is probably a low-mass WD system with a correspondingly low X-ray 
luminosity (Greiner 2000b) far below that observed for source N1.

\begin{center}
\includegraphics[angle=90,width=\columnwidth]{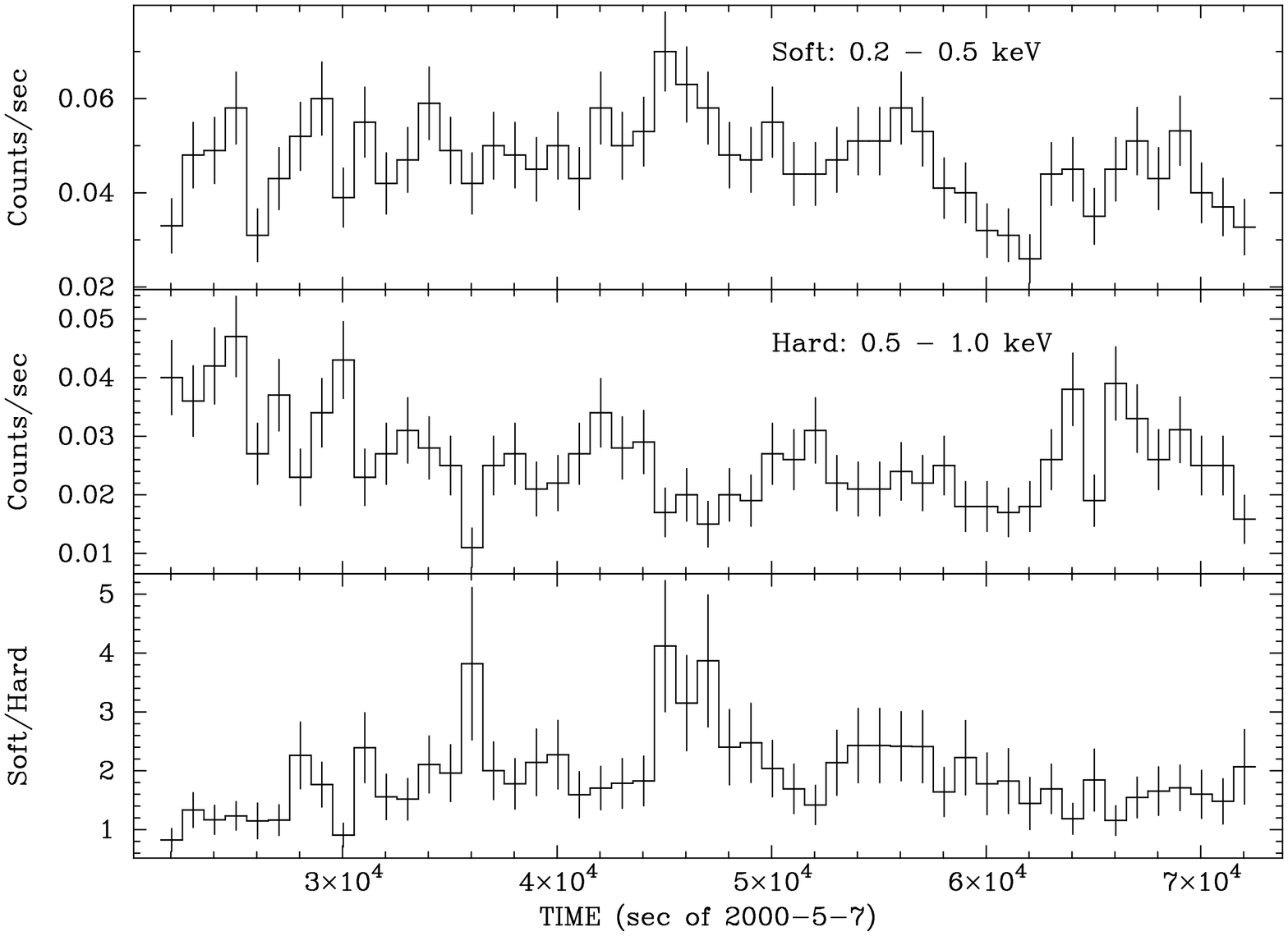}
\vspace{20pt}
\figcaption{\cha\ 50 ks light curve of source N1 in the 
0.2--0.5~keV ({\em top}) and the 0.5--1.0~keV ({\em middle}) energy bands 
binned on 1000~sec intervals.
The bottom panel displays the ratio of the two energy bands.
Error bars represent 1$\sigma$ statistical uncertainties.
  \label{f:2band_lc}}
\end{center}

Interestingly, the spectrum is substantially harder and the brightness
is higher during the bright portion of 
the 2.4~ks observation compared to the 50~ks observation.
There is even some flux above 1.0~keV in the 
2.4~ks data. 
The mean photon energy during this bright phase is $\sim$730~eV
while during the 50~ks observation the mean is $\sim$475~eV 
(Figure~\ref{f:sn1_spectralfit}).
Part of the spectral hardening, including most of the flux above
$\sim$1.1~keV, is caused by pileup in the high rate data. 
Nevertheless, the spectrum is definitely harder 
(or, hotter in the context of blackbody spectral models)
during this bright phase.
There is no clear
signature of increasing absorption as the source enters the low rate region
of the 2.4~ks light curve as would be expected in the onset of an eclipse. 
No other source shows a similar harder spectrum during the 2.4~ks observation.

The other alternative discussed previously 
in the context of spectral models for source N1 
is X-ray emission from an accretion disk around a mid-mass black hole. 
Variability on many timescales is observed in both 
massive black holes (AGNs) 
and in stellar-mass black holes (X-ray transients).
Variability is generally caused by instabilities in the disk structure
and changes in the accretion flow and is 
usually accompanied by spectral evolution among high soft
states and low hard states. The 50~ks light curve of source N1 in each of
two energy bands is shown in Figure~\ref{f:2band_lc} along with the ratio
of the soft band to the hard band. This shows that source N1 varies 
differently in different energy bands but it does not help distinguish between
standard \sss\ and mid-mass black hole scenarios. 

\subsection{Long-Term Variability}

The fluxes observed by \cha\ suggest the 
brightest 4 or 5 \sss\ candidates may be 
detectable in archival \ros\ observations of the field. 
M81 was observed 9 times by \ros/PSPC over the interval 
1991 March through 1994 April and 
11 times by \ros/HRI from 1992 Oct through 1998 April. 
Source N1 is present in \ros/HRI images but is too near 
the bright nucleus of M81 to be resolved in \ros/PSPC data. 
Source N2 is clearly present and isolated 
from other X-ray sources in both PSPC and HRI observations. 
Source N3 is confused with
a bright nearby object 
in both \ros\ instruments.
Source N4 is likewise too near the nucleus for positive detection. 
Source N5 is marginally
detected at the 2.8$\sigma$ level in the longest of the HRI exposures. 
This corresponds to a \cha\ luminosity of $\sim 2\times 10^{37}$~\ergl;
consistent with a constant source luminosity between the two 
observations.
The remaining sources
are too weak to be detected at a reasonable confidence level.
The entire \ros\ dataset has been independently analyzed by 
Immler \& Wang (2001). 
\cha\ source N1 is within $3.\arcsec 2$ of HRI source H25 of 
Immler \& Wang (2001) and is listed as 
a variable source with no other comment. 
Source N2 is within $3.\arcsec 3$ of HRI source H36
which is coincident with PSPC source P44. 
Source P44 is a soft source based on its PSPC hardness ratios 
(Immler \& Wang 2001).
No other \cha-detected \sss\ candidate is among the \ros\ sources identified by 
Immler \& Wang (2001).

The combined \ros/HRI, PSPC, and \cha\ light curves 
of sources N1 and N2 are displayed in 
Figure~\ref{f:rosat_lc}. Here, the average count rate 
of each individual observation is shown
combined to form each data point. The \cha\ and \ros/PSPC
observed count rates have been scaled to \ros/HRI observed count
rates using the PIMMS tool (Mukai 1993) and assuming spectral properties
for the two sources as derived in section~\ref{s:spectral_analysis}.
This scaling can account for some of the disparity seen, for example,
in the lower panel of Figure~\ref{f:rosat_lc} (source N2) where the
PSPC data and the two \cha\ points  
are lower than the intervening \ros/HRI fluxes. Nevertheless,
both sources are clearly present in the data spanning $>$7.5~yr. 
As discussed in the next section, this fact limits the possibility
that these sources could be classical novae.

\begin{center}
\includegraphics[angle=90,width=\columnwidth]{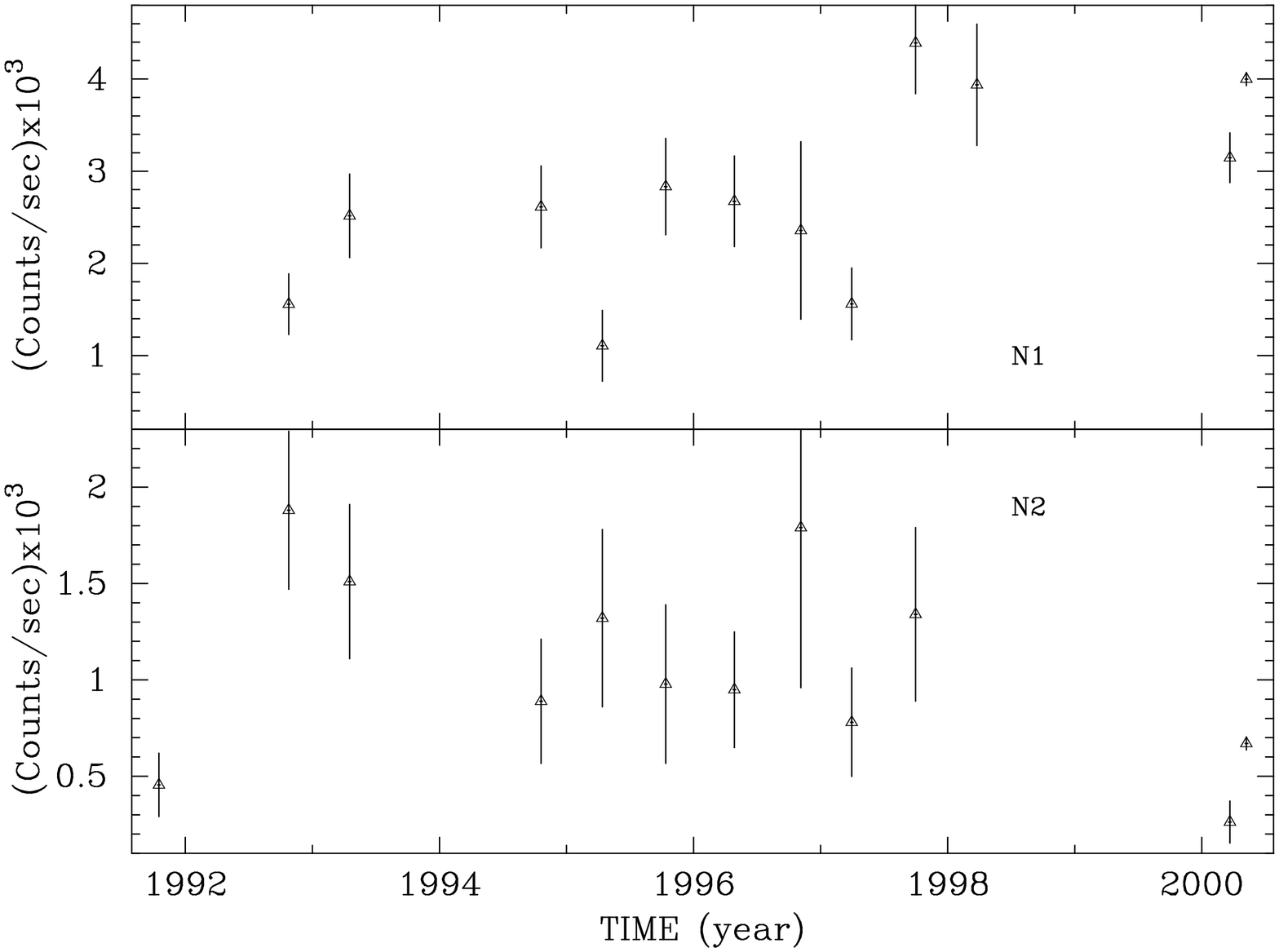}
\figcaption{Combined \ros\ and \cha\ light curves for 
Source N1 ({\em upper panel}) and N2 ({\em lower panel}). 
Count rates have been scaled to \ros/HRI count rates assuming the
spectral parameters derived in \S\ref{s:spectral_analysis} 
from the 50~ks \cha\ data. \ros/HRI data are marked by squares,
\ros/PSPC data with triangles, and \cha\ data with circles.
Errors denote 1$\sigma$ statistical uncertainties.
Labels denote average observed count rates. \label{f:rosat_lc}}
\end{center}

\section{Discussion} \label{s:discussion}

\Sss\ are a well-established class of X-ray emitting objects. 
Of the 19 well-studied systems (e.g., Greiner 2000a), 
all within the Galaxy or the Magellenic Clouds, 
nine are accreting WDs in close binaries (CBSS) 
that are nuclear burning accreting material in a steady state, 
four are classical or recurrent novae, 
three are symbiotic systems,
and one is a planetary nebula nucleus.
The common source of X-ray emission among these sources is nuclear burning 
either of accreting material (in a steady state or in a nova flash) 
or of residual fuel (following, {\em e.g.}, the formation of a planetary nebula).

The assumption that all \ssss\ result from 
nuclear burning on WD stars is adopted here for
purposes of discussion. Two questions to be addressed are: 
(1) Is the observed population and its distribution within M81 consistent with 
this theory and (2) Are the observed X-ray properties of the individual sources
consistent with previously-observed members of this rather heterogenous class
(and, if so, which ones)?

\subsection{The M81 Population of Supersoft Sources}

\begin{figure*}
\begin{center}
\hspace{30pt}
\includegraphics[angle=90,width=3.0in]{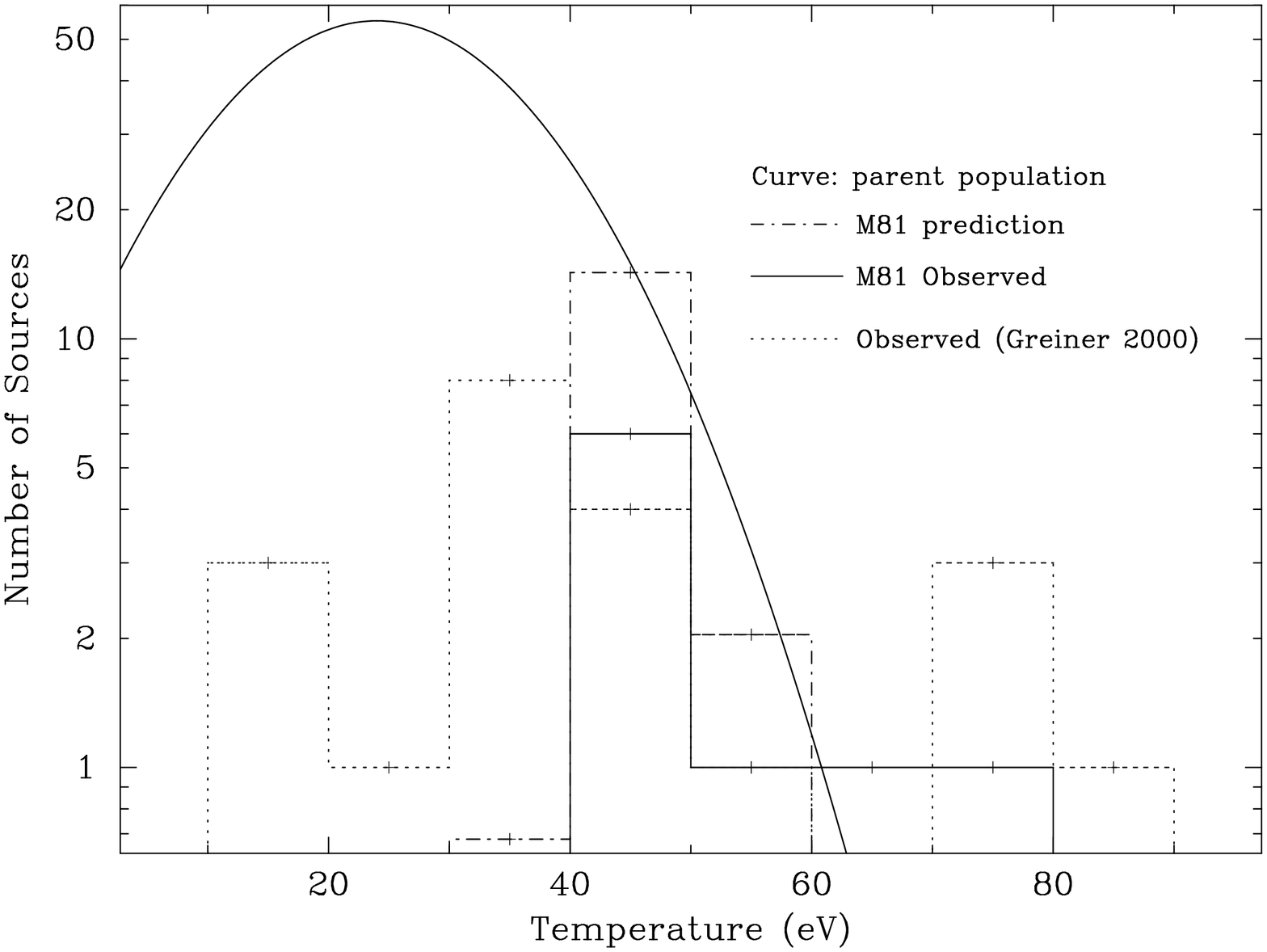}
\hfil 
\includegraphics[angle=90,width=3.0in]{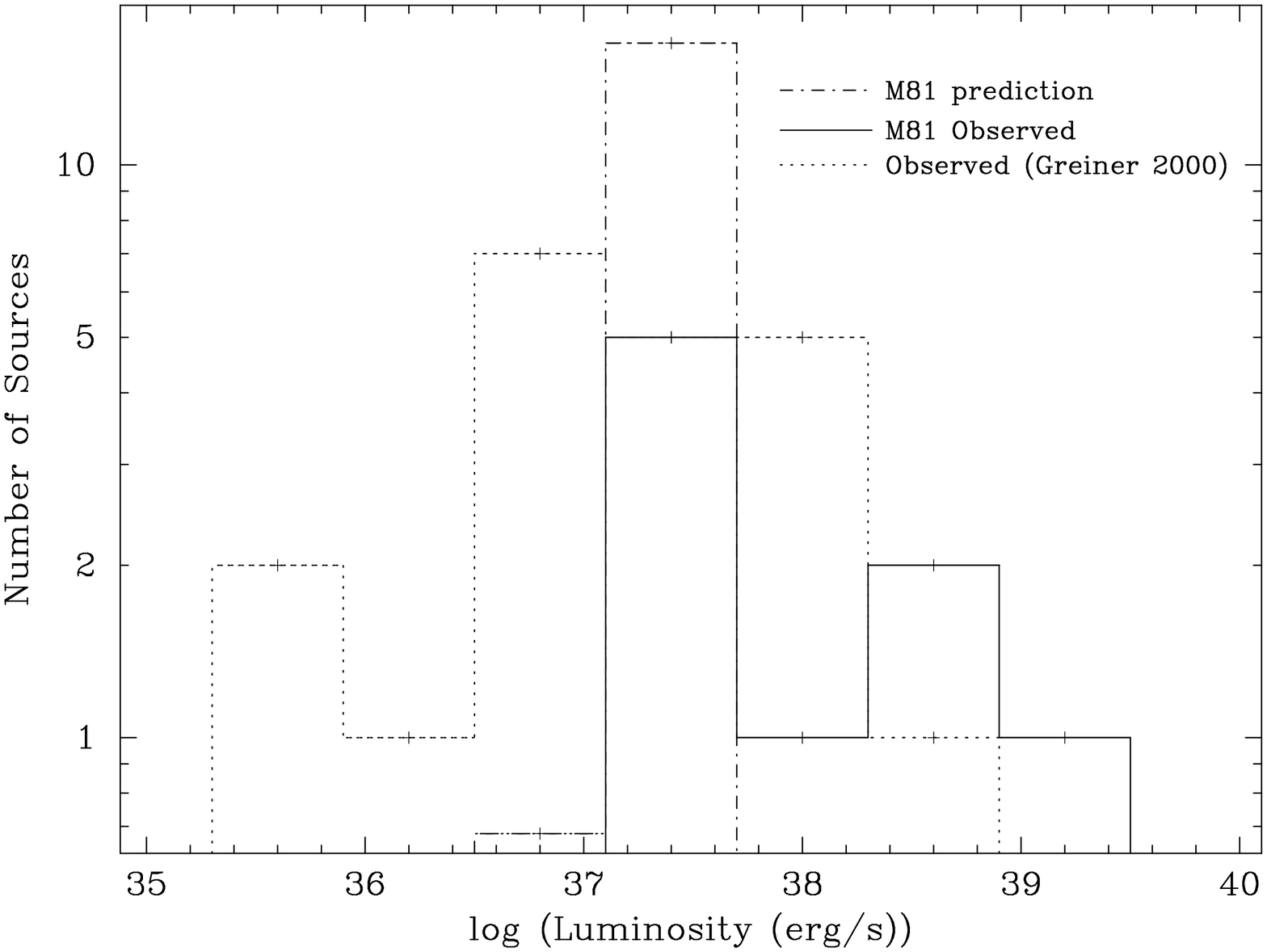}
\vspace{20pt}
\figcaption{Distribution of \sss\ systems over effective temperature 
({\em left}) and bolometric luminosity ({\em right}). 
Shown are the predicted and observed distributions for the galaxy M81 
(see text) and the distribution of known sources in the Galaxy and 
Magellenic Clouds as catalogued by Greiner (2000a). 
The M81 predicted distributions assume the sources
are distributed normally over both temperature and luminosity. 
This normal distribution is the curve in the left panel. For clarity,
the corresponding distribution is omitted from the right panel as it
appears as a narrow asymmetric curve
centered on 3$\times$10$^{37}$~\ergl\ in the logarithmic scaling used here.} 
\label{f:t_eff_l_bol}
\end{center}
\end{figure*}

Nine \sss\ candidates have been identified in M81 
based on broad-band X-ray colors.
All are located on the S3 device.
Four of these are located in the bulge of M81 
and the remainder are aligned with the 
spiral arms with the exception of one object in the inter-arm disk region. 
The bulge comprises 30\% of the area of the S3 
device\footnote[6]{The bulge is defined here as the region interior to
the inner Lindblad resonance at  
radius 4$\pm$0.2~kpc (Kaufman \etal\ 1989 and references therein). 
This is larger than the bulge radius (2.55~kpc) 
adopted by Tennant \etal\ (2001) 
resulting in 13 more X-ray sources within the bulge.}
thus 2.7 of the 9 \ssss\ should occur 
in the bulge if the spatial distribution were uniform on S3.
However, 54 of the 97 X-ray sources of all
types on S3 are located in the bulge (Tennant \etal\ 2001) 
and therefore 5 \ssss\ would be expected
there if the \ssss\ follow the general distribution of X-ray sources. 
Twenty-one of the 43 disk X-ray sources are aligned with the spiral arms
while 4 of the 5 supersoft disk sources are on the arms.

The observed properties of the \ssss\ in M81 favor high mass WD systems 
($M_{WD}$$\sim$1.0~\msun) and hence a younger population of progenitor stars 
(zero-age main sequence mass $\sim$7~\msun).
This favors locations near star forming regions such as spiral arms.
Conversely, 
the interstellar absorption is relatively higher in the spiral arms than in the
disk or bulge which would tend to obscure the lowest-temperature \ssss.
Indeed, the two \ssss\ located on the spiral arms 
with sufficient counts for spectral
fitting (N2 and N3) have relatively high X-ray absorbing columns 
and temperatures of 50--60~eV.

Objects in the bulge are expected to be an older population 
($>$9~Gyr, e.g., Kong \etal\ 2000)
except in the central $\sim$50$\arcsec$ where an enhancement in recent 
star formation activity is suggested by UV 
(Hill \etal\ 1992, Reichen \etal\ 1994) 
and H$\alpha$ (Devereux, Jacoby \& Ciardullo 1995) imaging
and consistent with the concept of gas-infall toward the center of M81.
This may have begun as recently as $\sim$400~Myr ago as a result of 
an encounter with NGC~3077 (Thomasson \& Donner 1993) or M82 
(de Grijs, O'Connell, Gallagher 2001).
All the bulge \sss\ candidates lie outside this central region
and appear to have relatively low absorbing columns consistent with the 
lack of gas and dust.
The standard \sss\ model (van~den~Heuvel \etal\ 1992) 
requires companion stars in the
mass range 1.3--2.5~\msun\ to maintain steady nuclear burning 
of accreting H on the surface of the WD. 
Stars of this mass range evolve from the main sequence on timescales of order 
several $10^8$ to greater than $10^9$ years 
and so the required accretion could be occuring in
the current epoch in the bulge if the donor stars were formed in an
encounter.

How many \ssss\ are expected in the 50-ks \cha\ observation? 
Di~Stefano \& Rappaport (1994) estimated the number of \ssss\ in external
galaxies observable by \ros\ by carefully accounting for 
the effects of interstellar absorption
on an assumed parent population taken from the work of 
Rappaport, Di~Stefano, \& Smith (1994). 
The parent population was that of systems 
whose luminosities and temperatures matched the 
steady-burning regime computed by
Iben (1982). This omits several classes of \sss\ 
including wind-driven symbiotic 
systems, novae, and recurrent novae. 
Nevertheless, it is possible to apply the basic approach of 
Di~Stefano \& Rappaport (1994) to
the \cha\ observation of M81. Scaling from their estimate of 
the total population
of \ssss\ in M31 by the ratio of blue luminosities
[{\em i.e.}, assuming \ssss\ contain massive WDs and hence evolved
from a young stellar population (Motch, Hasinger, \& Pietsch 1994,
Di~Stefano \& Rappaport 1994)],
a total of 1700 \ssss\ are expected in M81. 

To estimate the fraction detectable by \cha,
the population distributions in temperature and bolometric luminosity 
as derived by Rappaport, Di~Stefano, \& Smith (1994) can be 
roughly approximated by
normal distributions with means $T = 24$~eV and 
$L=3\times 10^{37}$~\ergl\ and standard deviations 
of 13~eV and $8.5\times 10^{36}$~\ergl, respectively 
(see also Fig.~3 of Di~Stefano \& Rappaport 1994). 
By randomly sampling values of $T$ and $L$ from these 
two independent distributions,
assuming a blackbody spectrum and a modest column of 
$N_H = 8 \times 10^{20}$~cm$^{-2}$ 
(twice the Galactic value), and using the PIMMS software tool (Mukai 1993), 
the predicted \cha\ observed count rate distribution can be accumulated. 
Assuming 20 counts in a 50-ks
observation as a detection threshold, 
only 1\% of the \ssss\ in M81 should be detectable with
the BI device S3. 

As the viewing field of S3 encompasses 
23\% of the $D_{25}$ area of
M81, $\sim$4 of the 1700 \ssss\ in the parent population 
should be detected in the 50-ks observation on S3. 
This is consistent with the observed population of 9 sources
in light of the large uncertainties involved including a 
factor-of-two uncertainty
in the normalization of the parent population 
(Di~Stefano \& Rappaport 1994) and the neglect of other classes of \sss\
systems by these authors (Yungleson \etal\ 1996).

The simulation also shows that only systems with temperatures $T\GA 40$~eV
and bolometric luminosities $\sim 2 \times 10^{37}$~\ergl\ 
are likely to be detected
by \cha. This result can be compared to the distribution of temperatures and
luminosities reported in Greiner (2000a) for 21 \ssss\ (two of which have not
been optically classified) and to the 9 \ssss\ discovered in M81.
Figure~11 shows this comparison. 
(For brevity, it has been assumed here that the 6 weakest \ssss\ in M81 all
have temperatures of 45~eV, see \S\ref{s:remaining_spectra}.)
The distributions of the Greiner (2000a) sample are clearly
broader than predicted or observed for M81. This is a consequence of
the greater diversity of local environments in which these sources are
found and the dissimilar instruments used to investigate them.  
The peak of the observed M81 distributions in temperature and luminosity 
coincide with the predicted distribution but these peaks are dominated
by the weak sources and are, therefore, highly uncertain. The bright observed
sources are, however, located at higher temperatures and luminosities 
than expected.

\subsection{The Nature of Individual M81 Supersoft Sources}

\subsubsection{Source N1}

By far the brightest \sss\ candidate in the sample 
lies within the bulge of M81 approximately 52$\arcsec$ 
from the nucleus. Its spectrum is well-fit with a simple
blackbody model of 78~--~86~eV ($T \sim 9 \times 10^5$~K) with 
a modest absorption ($N_H$$\sim$$6 \times 10^{20}$~cm$^{-2}$) 
resulting in an implied bolometric
luminosity exceeding $10^{39}$~\ergl. 
Although the bolometric luminosity is 
highly uncertain, even the observed luminosity exceeds
the Eddington limit for a Chandrasekar-mass WD.

The brightest well-studied \ssss\ are typically 
novae or recurrent novae with peak
bolometric luminosities often exceeding $10^{38}$~\ergl. 
One of the best examples is U~Sco. 
Kahabka \etal\ (1999) investigated BeppoSAX observations of 
this recurrent nova taken $\sim$20
days after the peak of the most recent optical outburst. 
They find a temperature of 
$\sim 9 \times 10^5$~K and a bolometric luminosity of up to 
$2 \times 10^{38}$~\ergl\ from a 
combination of non-LTE atmosphere models and an 
optically-thin thermal emission component,
assumed to arise from a wind, added to
fit the $\sim$1--2~keV portion of the observed spectrum.

Although source N1 has X-ray properties similar to those of U~Sco,
it is unlikely that source N1 is a nova. 
Novae occur in systems with relatively low mass transfer
rates where a layer of fuel is accumulated 
for a period of time before compressional
heating causes ignition. 
The mass of the accumulated layer increases with decreasing WD mass.
Consequently,
the X-ray light curves evolve slowly, lasting up to 10 years or so, 
in systems with low-mass WDs.   
The luminosity is proportional to the WD mass $M_{WD}$, 
$L_{bol}$$\sim$$1.8\times10^{38}(M_{WD}-0.26)$~\ergl\ assuming H accretion,
where $M_{WD}$ is measured in solar units (Iben \& Tutukov 1989). 
Thus the luminosity is low for these long-lived low $M_{WD}$ novae
and the recurrence time between events is very long. 
Conversely,
systems containing high mass WDs, 
such as inferred for U~Sco, have a higher peak luminosity 
but are short-lived X-ray sources ($\sim$0.1~yr, {\em e.g.}, Kato 1997). 
The recurrence times are also shorter but
are still of order years
and therefore inconsistent with the observed light curve of source N1.
(Note the short- and long-term variations observed in the 
X-ray light curve of source N1 are better described as
transience or variability in contrast to recurrence
as applied to novae.)

At higher accretion rates is a regime of steady-state
burning of accreting material.
The maximum luminosity in the steady-burning regime is 
$L_{bol} = 2.3\times10^{38}(M_{WD}-0.5)$~\ergl\
and occurs at the maximum accretion rate 
\mdot$=7.8\times10^{-7}(M_{WD}-0.5)$~\msun-yr$^{-1}$ 
(Iben 1982, again assuming H-rich accretion). The 
highest temperatures, up to $\sim$85~eV, 
are obtained by the highest mass WDs 
as are the highest luminosities. 
However, an accretion rate of $\sim 5\times10^{-6}$~\msun yr$^{-1}$ is
needed to obtain the bolometric luminosity inferred for source N1 
assuming steady burning.
At such high accretion rates, however, the photospheric radius 
expands to red giant dimensions;
only a fraction of the fuel is consumed; and the X-ray
emission, formed in a thin layer at the base of the envelope, 
is hidden by the overlying material.
The burning may instead drive a strong, opaque wind  
(Hachisu, Kato \& Nomoto 1996).
If the outcome of a high mass transfer rate is formation of an envelope 
then this system 
should be observable optically as a hot giant or red giant star, 
contrary to the \hst\ observations 
(\S\ref{s:environments}) and not as a strong X-ray source.
If the accretion drives a wind, 
then the photospheric temperature is only $T\sim 10^5$~K.
However, if the accretion rate {\em decreases} 
the wind mass-loss rate will decrease as will 
the effective photospheric radius. 
Regions closer to the burning layer will become
exposed and the X-ray temperature and X-ray luminosity 
will increase dramatically (Hachisu, Kato, \& Nomoto 1996). 
The subsequent evolution of systems of this type
were not followed by Hachisu, Kato, \& Nomoto (1996).

If the accreting material is He-rich, then the system can sustain a
higher mass transfer rate while remaining in the steady nuclear-burning 
regime (Iben \& Tutukov 1989). The increase in mass transfer rate by roughly
an order of magnitude translates into only a factor of 2 or so higher 
luminosity because of the lower specific energy generation rate of He 
relative to H so that the maximum luminosity is
$L_{bol}$$\sim$$4.6\times10^{38}(M_{WD}-0.58)$~\ergl
(Iben \& Tutukov 1989). 
Although He-rich LTE white dwarf atmosphere models failed to
provide a satisfactory fit to the observed spectrum,
He-rich acccretion remains an attractive scenario for source N1 because
the Eddington limit luminosity for He accretion
is twice that for H. Helium accretion can occur during a second phase of 
mass transfer to the compact star when the companion has exhausted He in
its core (Iben \& Tutukov 1994). The donor star in this scenario is
either the nucleus of an early asymptotic giant branch or an evolved 
He-star remnant with a CO core.
Under appropriate conditions, this mass transfer can be conservative
(Iben \& Livio 1993) instead of lost in a common-envelope process (Iben \& Tutukov 1985). The initial mass of the donor star tends to be rather large
in this scenario, $\sim$6.5--9.5~\msun\ for steady burning, 
so that such systems should be rather
rare in the old population of bulge stars in M81. 
Helium tends to burn explosively (Sion \& Starrfield 1993) unless
the helium layer can be maintained at a high temperature (Jos\`{e} \etal 1993).
A series of weak He shell flashes may occur for systems
near the stability limit producing luminosities of order 
$\sim$$4\times 10^{38}$~\ergl, and temperatures $\sim$70~eV 
(Iben \& Tutukov 1994).
These systems tend to have less than $\sim$20~hr orbital periods
(Yungelson \etal\ 1996),
comparable to or less than those of subgiant donors in H-accreting systems 
and of main sequence stars 
losing mass on a thermal timescale ($\sim$1day orbital periods,
van~den~Heuvel \etal\ 1992).
Another channel for He accretion is a semidetached double degenerate system
in which the secondary is a helium WD (Iben \& Tutukov 1989). 
Such a system must have an orbital period of order minutes 
(Verbunt \& van~den~Heuvel 1995).

In short, none of the standard nuclear-burning WD scenarios predict
luminosities as high as that inferred for source N1. 
However, theoretical treatments have not rigorously explored the behavior
of these systems in regimes beyond the stability limit.
Other alternatives exist. Neutron star binaries often exhibit a luminous
thermal component though rarely as soft as observed here (and hence one of 
the original motivations for the nuclear-burning WD sceanrio for \ssss, 
van~den~Heuvel \etal\ 1992). One exception is
the Be/X-ray binary RX~J0059.2-7138 with $T_{eff}$$\sim$36~eV (Hughes 1994)
though this source is accompanied by a strong power law component extending 
the spectrum beyond the \ros\ band. Weakly magnetized neutron stars such as
the X-ray burst sources, tend to have correspondingly weak power law 
spectral components but also have high characteristic temperatures during 
the burst phase (e.g., Lewin, van~Paradijs, \& Taam 1995). Similarly, typical
black hole binaries display a ``soft'' component with characteristic 
temperature $>$1~keV and a hard power law component.

A completely different scenario that may explain the observed X-ray 
properties of source N1 is X-ray emission from an optically-thick 
accretion disk surrounding a mid-mass black hole. The most attractive
feature of this model is that the implied bolometric luminosities 
derived here are 
a small fraction of the Eddington limit. The observed X-ray spectrum is 
consistent with that from the innermost disk radius assuming the central
object has a mass $M_{BH}$$\sim$1200/(cos$i)^{1/2}$. While no objects 
in this mass range have been confirmed, invoking the Eddington limit
to X-ray-bright sources has led to suggestions of masses 
in excess of 10-100 \msun, in many instances ({\em e.g.}, Makishima \etal\ 2000) and as high as $\sim$700 \msun\ in the case of a luminous source
in M82 (Kaaret \etal\ 2001, Matsumoto \etal\ 2001).

\subsubsection{Source N2}

Source N2 has a distinct edge at $\sim$870~eV
well fit by O~VIII absorption in the LTE atmosphere model. 
This feature is commonly found in the hotter \ssss\ such as the LMC 
source CAL~87. 
Ebisawa \etal\ (2001) find a temperature of 75~eV for CAL87 from
\asca\ observations with an absorbing column comparable 
to that obtained here for source N2 ($N_H \sim 2 \times 10^{21}$~cm$^{-2}$). 

Parmar \etal\ (1997) report 
BeppoSAX observations of CAL~87  are equally well-fit
with either blackbody, LTE, or NLTE atmosphere models with temperatures
ranging from 42 eV for the blackbody fits to 57 and 75~eV for their LTE
and NLTE models. The blackbody model implies  
$L_{bol}$$\sim$$4\times 10^{38}$~\ergl\ but 
the LTE and NLTE fits are consistent with more modest luminosities,
$\sim$3--$5 \times 10^{36}$ \ergl. 
Including the O VIII absorption edge increases the blackbody temperature
and decreases the bolometric luminosity to values similar to their NLTE models.

This trend is opposite to the results reported here for source N2: blackbody
models predict higher temperatures than the LTE models and bolometric
luminosities comparable to or lower than the LTE models. The best-fit model,
LTE with 1\% of solar metal abundance, 
has a temperature of 67~eV and a hydrogen
column density of $1.6 \times 10^{21}$~cm$^{-2}$ very similar to CAL~87.
The observed luminosity is $3.9\times 10^{37}$~\ergl\ and
$L_{bol}$$\sim$$5.3^{+4.8}_{-2.1}\times 10^{38}$~\ergl, 
much higher than the models of 
Parmar \etal\ (1997) predict but consistent (within the
large uncertainties) with the highest
luminosity reported by Ebisawa \etal\ (2001) for CAL~87.

CAL~87 is a close binary system that exhibits eclipses with a period of 10.6 hours. The eclipses are visible as shallow dips in the X-ray light curve. 
Small-amplitude variability of this kind 
cannot be distinguished in the relatively low-quality
light curve of source N2.

\subsubsection{Source N3}

This source is also relatively highly absorbed but considerably cooler than the 
two brightest sources with $T_{eff} \sim 50$~eV. 
This value is more typical of many of the well-studied \ssss\ 
(Figure~\ref{f:t_eff_l_bol}, Greiner 1996, 2000a), such as the CBSS sources
CAL~83 and 1E~0035.4-7230. 
Parmer \etal\ (1998) report acceptable fits can be achieved with blackbody,
LTE, or NLTE atmosphere models applied to BeppoSAX observations of CAL~83.
Their resulting 
temperatures are $\sim$45~eV for the blackbody and LTE models and $\sim$33~eV
for their NLTE fits.  
As with source N3 (\S\ref{s:sn3}), Parmer \etal\ (1998) report 
the low temperature and consequent 
lack of strong absorption edges does not allow a distinction to be made 
among the models (see also Figure~\ref{f:lte_models}). 
BeppoSAX and \ros\ observations
of 1E~0035.4-7230 (Kahabka, Parmer, \& Hartmann 1999) again do not allow
either blackbody, LTE, or NLTE models to be discounted though in this case
a feature ascribed to C~V and C~VI absorption is evident. Temperatures 
derived by Kahabka, Parmer, \& Hartmann (1999) range from $\sim$40$\pm$13eV
from their blackbody fits to $\sim$28$\pm$4~eV from their NLTE model fits to 
1E~0035.4-7230. Both CAL~83 and 1E~0035.4-7230 derived 
bolometric luminosities are of order a few times $10^{37}$, roughly an order
of magnitude lower than the best-fit values derived here for source N3 though
within the large uncertainties in the latter.

\subsubsection{Source N4 and N5} \label{s:s4ands5}

Though little can be said about the X-ray properties of these two 
sources beyond their broad-band classification as \sss\ candidates, 
these objects may be similar to the LMC source CAL~83.

Source N4 and N5 are spatially coincident with objects ID~68 and 116,
respectively, in the list of [OIII] $\lambda$5007~\AA\ sources 
observed by Jacoby \etal\ (1989). This emission line is also prominent
in the optical images of CAL~83 (Remillard \etal 1995) and is
attributed there to ionization in the interstellar medium by 
a large UV and soft X-ray photon flux from the central source.
Rappaport \etal\ (1994) modeled the ionized regions surrounding \ssss\ 
and calculated resulting optical line intensities.
The strongest predicted lines 
are the [OIII] $\lambda$5007~\AA\ and HeII~$\lambda$4686~\AA\ lines. 
Adopting the foreground extinction to M81 from Jacoby \etal\ (1989), 
their observed [OIII] $\lambda$5007~\AA\ fluxes are consistent 
with those expected from the \sss\ models. 
For example, the $\lambda$5007~\AA\ flux from
object ID~116 is $\sim$$4 \times 10^{-16}$~erg~cm$^{-2}$~s$^{-1}$ 
while Rappaport \etal\ (1994)
predict $1.6 \times 10^{-16}$~erg~cm$^{-2}$~s$^{-1}$ and 
$2.3 \times 10^{-15}$~erg~cm$^{-2}$~s$^{-1}$
from models with intrinsic luminosities of $10^{37}$ 
and $10^{38}$~erg~s$^{-1}$,
respectively, and source temperatures of $4 \times 10^5$K.

Parmar \etal\ (1998) find a temperature $T_{eff}$$\sim$30-50~eV for CAL~83
and $L_{bol}$$\sim$$4 \times 10^{37}$~\ergl. These values are
within the estimated values for sources N4 and N5.

A PNe cannot be excluded based on the X-ray and [OIII] fluxes of sources N4 
and N5. The central stars of PNe can have luminosities as high as
that inferred for sources N4 and N5 but only the most massive central stars
are hot enough to be detected in soft X-rays (e.g., Paczynski 1971). This is 
the case for the SMC \sss\ 1E00056.8-7154. This source is 
coincident with PNe N67 (Wang 1991) but is anomalously X-ray bright for a PNe 
[$T_{eff}$$\sim$38~eV $L_{bol}$$\sim$$2\times 10^{37}$~\ergl\ 
(Brown \etal\ 1994, Heise \etal\ 1994)]. Unfortunately, no obvious 
spectroscopic distinction between PNe and \ssss\ has been identified
(Di~Stefano, Paerels, \& Rappaport 1995).

\subsubsection{Other Sources}

The X-ray evidence in support of a \sss\ nature of the 
remaining candidates is based 
entirely on their X-ray colors. 
Their spectra and light curves are of insufficient 
quality for detailed analysis. 

As there is no long-term X-ray light curve for these sources, their
identification as novae cannot be ruled out. The rate of nova
occurrence in M81 is rougly 20~yr$^{-1}$ (Della~Valle \& Livio 1994)
and, as discussed above, 
X-ray lifetimes range from months for the brightest sources up
to $\sim$10 years. 
For the values of $L_{bol}$ estimated in \S\ref{s:spectral_analysis}, 
$10^{37}$ to
$4\times 10^{37}$~\ergl, the corresponding WD masses are low,
$M_{WD} \sim 0.3$ to 0.5~\msun. These masses correspond
to long-lived novae, typically radiating as a (very soft) 
X-ray source for $\sim$7~yr for $M_{WD}$$=$0.6~\msun\ (Kato 1997). 
Visual magnitudes of novae  
can be as high as $m_V \sim 22$ to 23 at the 
distance of M81 but the visible light rapidly declines after
maximum light as the spectrum hardens and the X-ray flux rises.
A 0.6~\msun\ WD nova, for instance, only becomes X-ray visible
$\sim$4~yr after optical outburst.

The weak \sss\ candidates could also be steady-burning CBSS systems.
The estimated $L_{bol}$ corresponds to a range of WD masses $M_{WD}$$\sim$0.7
to $\LA1.0$. However, WDs in this mass range should be considerably
cooler ($T_{eff}\LA\ 40$~eV) than the 40--50~eV temperatures suggested by the 
instrumental selection effect. Below $M_{WD}$$\sim$0.65~\msun, 
the photospheric radius expands in the steady nuclear burning case
and $T_{eff}$ plummets (Iben 1982). Thus only a narrow
range of masses are possible for the weak \ssss\ if they are steadily
burning their accreted hydrogen.

Another evolutionary channel leading to the formation of a \sss\ is
wind-driven accretion in symbiotic systems
(Sion \& Starrfield 1994). These systems can undergo H shell
flashes, reach luminosities of order $\sim$$10^{37}$~\ergl, and remain
at this plateau for $\sim$250 yr in some circumstatances. 
For many of the low-mass WDs modeled by Sion \& Starrfield (1994), 
the flashes are weak, the envelope radius remains
compact, and the resulting photospheric temperature
is relatively high. However, the maximum temperatures achieved
were $T_{eff}$$\sim$$20$~eV. 
The \sss\ SMC~3 (RX~J0048.4-7332) is a symbiotic nova of this type.
Jordan, \etal\ (1996) derive a luminosity 
$L_{bol}$$\sim$$4\times 10^{37}$~\ergl, a temperature of 22~eV, and
a mass of $\sim$0.8~\msun\ for this object.
The weak \sss\ candidates are unlikely to be symbiotic systems
because of the difficulty detecting objects of such low temperatures
in M81 with \cha. 
An archival \cha\ observation of AG Dra, for 
instance, detects very little flux above $\sim 0.3$~keV. 

\section{Summary \& Prospects} \label{s:conclusions}

\subsection{\cha\ Observations of M81}

An important part of the \ros\ legacy has been the identification 
of \ssss\ as a phenomenologically distinct class of X-ray emitting object.
The \cha\ X-ray Observatory affords an excellent opportunity to extend the 
study of \ssss\ to other nearby galaxies.

As a first step, the population of \sss\ candidates discovered in M81
has been investigated. 
The number of \ssss\ detected and their spatial distribution are
consistent with population synthesis estimates.
X-ray spectral analysis shows the M81 sources
are qualitatively similar to previously-studied \ssss\
though the brightest M81 sources
tend to be hotter and more luminous than is typical of the class.
This was shown to be, in part, 
a natural consequence of the higher sensitivity
of \cha\ to the high-temperature region of the \sss\ distribution.

Bolometric luminosities deduced here from the spectral fits often exceed
the Eddington limit for spherical accretion onto a 1.4 \msun\ star.
Such extreme luminosities lie beyond the previously-observed
distribution of \ssss\ and, more importantly, cannot be reconciled 
with prevailing theories for \ssss\ as 
WDs powered by stable, steady-state, surface nuclear burning.
But,
the study of \ssss\ with the \cha\ Observatory is still in its infancy.
Important uncertainties remain. As in the \ros\ era,
spectral modeling remains problematic. Simple blackbody models, like those
used in this work, are known to greatly overestimate bolometric luminosities
in some cases. LTE model atmospheres, also applied here, are appropriate when a
well-defined photosphere is present such as in the high density gradient
at the surface of a WD. Even this model fails when the nuclear burning
is near the stability limit, the atmosphere becomes extended, and a wind
drives mass loss as may be the case in the brightest \sss\ candidate in M81.
In addition, the energy calibration of the BI devices at low photon energies
is uncertain and could also have a significant effect on the results
reported here.

Many of these uncertainties may be resolved or at least mitigated
as pending \cha\ observations of well-studied \ssss\ are analyzed and
compared to previous observations and to specific theoretical expectations.
High-resolution grating spectroscopy will be
indispensible in this regard as many of the calibration uncertainties
can be circumvented while, at the same time, models for the X-ray
spectra can be more tightly constrained.
An example of the richness of such a spectrum is 
that of CAL~83 obtained by XMM-Newton/RGS (Paerels, \etal\ 2001).
The population of \sss\ candidates discovered in M81 also
reveals the value of isolating sources to investigate their
local environments and of obtaining 
precise locations for future observations
at X-ray and other wavelengths.

In spite of the current uncertainties, the
remarkable properties deduced for the brightest \sss\ candidate 
in M81 should not be overlooked. 
This source exhibits large variations in X-ray brightness on timescales
of order hours yet maintains a high average X-ray brightness
throughout the roughly 8 year span of combined \ros\ and \cha\
observations of the region. 
In the context of the nuclear burning WD scenario, 
its high temperature and luminosity imply a high mass, near the Chandrasekhar
limit for WDs, and a high accretion rate, above that for 
steady-state nuclear burning.
While optical monitoring would be difficult, at best, future
observations 
are strongly encouraged to more fully understand this source. 
The accurate location provided in Table~1 should aid substantially
in this regard.

\subsection{Observations of Other Galaxies}

The most extensive study of \ssss\ in an external spiral galaxy similar to our
own and M81 is the \ros\ survey of M31 (Supper \etal\ 1997, Kahabka 1999).
Based on hardness ratios, Supper \etal (1997) identified 15 \sss\ candidates
not associated with SNRs or foreground objects in a 6.3 square degree M31 
field.
Kahabka (1999), using a slightly modified selection criteria, identified
an additional 26 candidates in the field. Kahabka (1999) found a total of
7 of the combined 41 \ssss\ are within the $\sim$6~kpc bulge of M31
with the remainder evenly distributed over $\sim$12 to 25~kpc radii. The
lack of detected sources in the region $\sim$6 to 12~kpc was explained as a 
consequence of a higher H column within this annulus. 
Kahabka (1999) suggests that the spatial distribution favors a disk
population of younger stars by comparing to the populations of Cepheids 
and (older) blue stars. The \sss\ distribution does not follow that of 
novae which are bulge-dominated and follow the old stellar population
found in the bulge. After accounting for possible foreground objects and SNRs,
Kahabka (1999) finds 1 bulge source for every 4--7 disk sources. This is 
marginally less than reported here for M81 where 4 bulge sources were found for
an extrapolated total 
population of 9/0.57$\sim$16 \ssss\ within the $D_{25}$ area of M81.
Kahabka (1999) also estimates the 
temperatures for all the sources and finds they range from   
$\sim$30--60~eV, typically, up to $\sim$73~eV for the recurrent
transient RX~J0045.4+4154. This latter
source is also one of the brightest with an inferred 
$L_{bol}$$\sim$$10^{38}$~\ergl\ (White \etal\ 1995).

A more detailed comparison between the \sss\ populations of M31 and M81
must await further analysis of the extensive \cha\ M31 dataset. Only then 
can many of the instrumental selection effects be eliminated. In the interim,
comparison to other \cha-observed galaxies must suffice.

Numerous deep observations of nearby galaxies have been performed by \cha\ in the $\sim$2 years since its launch. Discovery (or lack thereof)
of \sss\ candidates have been reported in the literature
for only a few galaxies.  
Sarazin, Irwin, \& Bregman (2000,2001) report 3 
\ssss\ in a sample of 90 point sources 
in the elliptical galaxy NGC~4697 to a limiting luminosity of 
$\sim 5 \times 10^{37}$~erg~s$^{-1}$ based on hardness ratios.
No \ssss\ were discovered in M84, 
an elliptical galaxy in the core of the Virgo cluster, 
based on the hardness ratio criteria (Finoguenov, Jones, \& Kudritzki 2001).
Pence \etal\ (2001) report the discovery of ten supersoft 
sources among 110 objects in M101 
using color definitions different than those defined here. 
Pence \etal\ (2001) combined 
the spectra of the three most luminous \ssss, 
reportedly having similar spectral shapes, and find a 
best fit blackbody temperature of 72$\pm$2~eV and a mean
(unabsorbed) luminosity of $1.4\times 10^{38}$~\ergl.
These values are similar to the brightest \ssss\ in M81 with the 
exception of source N1 which has an unabsorbed flux 5--8 times higher. 
The remaining 7 \ssss\ in M101 have a distinctly softer spectrum.
When combined, the best fit blackbody temperature is 47$\pm$2~eV 
and unabsorbed luminosity is $L_{X}=1.1\times 10^{37}$~\ergl\
for a column fixed to the Galactic value 
($N_H = 1.2 \times 10^{20}$~cm$^{-2}$).
Again, these values are consistent with those estimated here for the
weak sources in M81 although, as shown above, a temperature of order
40--50~eV is expected on the basis of instrumental selection effects
and the shape of the parent population estimated from population synthesis
models.
The position of only one \sss\ candidate lies within the bulge of M101.
However, M101 has a later Hubble type, Scd, and 
hence a smaller bulge than M81 (Sab)
extending only to $\sim$0.$\arcmin$75, or approximately 3\% of the S3 viewing
field analyzed by Pence \etal\ (2001).

These comparisons are intriguing. They suggest that perhaps there is
another \sss\ population of anomolously X-ray bright and hot objects in
nearby galaxies similar to our own. Care must be exercised, of course,
as there are, to date, only a handful of objects with these properties, 
no corroborating evidence in support of this conjecture from other
wavebands, and no satisfatory theory predicting such extreme X-ray behavior.
These comparisons 
also suggest a trend across the Hubble sequence with fewer \ssss\ in
early type galaxies compared to late types. 
The \ssss\ appear to be 
associated with the younger population of stars found on spiral arms though 
some are found in the relatively old population of bulge stars.
Irregular galaxies would then
have a disproportionate number of \ssss. 
Di~Stefano \& Rappaport (1994) estimate 125 and 25 \ssss\ are active in the 
LMC and SMC, respectively, based on their blue luminosities relative to M31.
Eight \ssss\ have been observed in the LMC and 4 in the SMC. 
This suggests either the population estimate is low or the observed fraction
is high for these two galaxies. Of course, there is a strong selection bias 
that must be taken into account. 

\vspace{0.125in}
We thank A. Ibragimov for constructing the WD model
atmospheres used in this work, A. Shafter for providing an H$\alpha$ image
of the field and for information on nova rates, T. Pannuti for 
radio data analysis, and M. McCollough for analyzing archival \hst/WFPC2 data.
K.~W.  and V.~S. thank M. Weisskopf for funding their visits to MSFC.
Support for this research was provided in part by NASA/\cha\ grant GO0-1058X
to D.~A.~S. and by Russian Basic Research Foundation grant 99-02-17488 to V.~S.

\end{document}